\pdfoutput=1
\PassOptionsToPackage{pdftex,
pdfencoding=auto,
pdfnewwindow=true,
pdfusetitle=true,
bookmarks=true,
bookmarksnumbered=true,
bookmarksopen=true,
pdfpagemode=UseThumbs,
bookmarksopenlevel=1,
pdfpagelabels=false,
breaklinks=true,
}{hyperref}
\PassOptionsToPackage{usenames,dvipsnames,pdftex}{xcolor}
\documentclass[12pt,letterpaper,onecolumn,final]{belarticle2}
\usepackage[square,sort,comma,numbers]{natbib}
\usepackage[showerrors,immediate]{silence}
\usepackage{amsmath}
\usepackage{amssymb}
\usepackage{epsfig}
\usepackage{textcomp}
\usepackage{amsthm, bbm}
\usepackage{mathrsfs}
\usepackage{stackrel}
\usepackage{subfigure}
\usepackage{cancel}
\usepackage{tikz}
\usepackage{paralist}
\usepackage[matha,mathb]{mathabx} 
\usepackage{accents}
\usepackage{mathtools}

\usepackage[usenames,dvipsnames]{xcolor}
\definecolor{purple}{RGB}{128,0,128}
\definecolor{ultramarine}{RGB}{63, 0, 255}
\definecolor{medblue}{RGB}{0, 0, 100}
\definecolor{googleblue}{RGB}{34, 0, 204}
\definecolor{panblue}{RGB}{0,24,150}
\definecolor{carmine}{RGB}{150, 0, 24}
\definecolor{gray}{RGB}{150, 150, 150}
\newcommand{\term}[1]{\textcolor{medblue}{\textbf{#1}}}
\AtBeginDocument{
\usepackage{hyperref}
\hypersetup{
linkcolor=carmine,
citecolor=googleblue,
urlcolor=panblue,
anchorcolor=OliveGreen,
}}

\usetikzlibrary{calc,decorations.pathreplacing}
\usetikzlibrary{arrows,shapes}

\newcommand{\cC}{\mathcal{C}}
\newcommand{\cB}{\mathcal{B}}
\newcommand{\cH}{\mathcal{H}}

\newcommand{\cQ}{\mathcal{Q}}
\newcommand{\cG}{\mathcal{G}}
\newcommand{\overbar}[1]{\mkern 1.5mu\overline{\mkern-1.5mu#1\mkern-1.5mu}\mkern 1.5mu}
\newcommand{\defin}{:=}
\newcommand{\disj}{\mathrm{disj}}
\newcommand{\comm}{\mathrm{comm}}
\newcommand{\FR}{\mathrm{FR}}

\let\OLDthebibliography\thebibliography
\renewcommand\thebibliography[1]{
  \OLDthebibliography{#1}
  \setlength{\parskip}{0pt}
  \setlength{\itemsep}{0pt plus 0.3ex}
}

\newtheorem{theo}{Theorem}

\newtheorem{prop}[theo]{Proposition}

\newtheorem{conj}{Conjecture}
\newtheorem{defn}{Definition}

\theoremstyle{definition} 

\newcommand{\OG}{\ensuremath{\operatorname{OrthoGraph}}}
\newcommand{\NOG}{\ensuremath{\operatorname{NonOrthoGraph}}}
\begin{document}
\title{Multipartite Composition \\of Contextuality Scenarios}
\author{Ana Bel{\'e}n Sainz and Elie Wolfe}
\affiliation{Perimeter Institute for Theoretical Physics, 31 Caroline St. N, Waterloo, Ontario, Canada, N2L 2Y5.}
\date{\today}

\maketitle

\begin{abstract}
Contextuality is a particular quantum phenomenon that has no analogue in classical probability theory. Given two independent systems, a natural question is how to represent such a situation as a single test space. In other words, how {separate} contextuality scenarios combine into a joint scenario. Under the premise that the the allowed probabilistic models satisfy the No Signalling principle, Foulis and Randall defined the unique possible way to compose two contextuality scenarios. When composing strictly-more than two test spaces, however, a variety of possible composition methods have been conceived. Nevertheless, all these formally-distinct composition methods appear to give rise to observationally equivalent scenarios, in the sense that the different compositions all allow precisely the same sets of classical and quantum probabilistic models. This raises the question of whether this property of invariance-under-composition-method is special to classical and quantum probabilistic models, or if  {it} generalizes to other probabilistic models as well, our particular focus being $\cQ_1$ models. $\cQ_1$ models are physically important  {since,}  when applied to scenarios constructed by a particular composition rule, coincide with the well-defined ``Almost Quantum Correlations". In this work we see that some composition rules, however, give rise to scenarios with inequivalent allowed sets of $\cQ_1$ models. We find that the non-trivial dependence of $\cQ_1$ models  {on} the choice of composition method is apparently an artifact of failure of those composition rules to capture {the orthogonality relations given by the Local Orthogonality principle}. We prove that $\cQ_1$ models satisfy invariance-under-composition-method for all the constructive compositions protocols which do capture this notion of Local Orthogonality.
\end{abstract}

\section*{Introduction}

Contextuality is a phenomenon that starkly reveals the difference between quantum and classical mechanics. 
Indeed, Kochen and Specker \cite{KS} proved that quantum mechanics is incompatible with the assumption that measurement outcomes are predetermined by physical properties independent of the context being measured. In this sense, classical theory fails at reproducing the measurement statistics obtained from operating on certain quantum systems. The original work by Kochen and Specker opened the door to different formalisations of contextuality, such as the sheaf-theoretic approach pioneered by Abramsky and Brandenburger \cite{AB}, the hypergraph-theoretic approach by Ac\'in, Fritz, Leverrier and Sainz \cite{AFLS}, the graph-theoretic approach of Cabello, Severini and Winter \cite{CSW}, and Spekkens' work on measurement and preparation contextuality \cite{SPE1, SPE2}, among others. 

In this work we will consider the phenomenon of contextuality as defined in Ref. \cite{AFLS}, where a contextuality scenario (also known as test space \cite{ss1,ss2}) refers to a specification of a collection of measurements which says how many possible outcomes each measurement has and which measurements have which outcomes in common. A contextuality scenario is hence defined by a hypergraph.
This allows us to study both the jointly-measurable observables approach \cite{AB} and the phenomenon of nonlocality \cite{BELL} in a unified manner within a general family of scenarios.  

A primary question in this framework is that of composition, i.e. given independent contextuality scenarios, how to define the hypergraph that represents the joint state space. We here consider only those composition methods which {at-minimum} ensure that the allowed correlations in the joint hypergraph respect the No Signalling principle. For the case of two scenarios, Foulis and Randall \cite{FR} were the first to propose a composition rule that satisfies such properties. For multipartite scenarios, however, there is more than one way to compose the individual hypergraphs in order to capture No Signalling \cite{AFLS}, hence forming a family of composition methods. A curious property of the entire family is that the sets of both classical and quantum correlations allowed in a {composite} scenario are independent of the specific choice of composition method. A natural question then is whether this property of invariance-relative-to-composition-rule holds for all quantifiable sets of correlations, or if it is a particular feature of the classical set and quantum set. 

A particular set of correlations called $\cQ_1$ (see Def.~\ref{q1mdef}) has caught the attention of the community, since these correlations satisfy many quantum-like properties \cite{AQ, AFLS}. Here, we ask whether the set of $\cQ_1$ correlations is also invariant under all composition methods which capture No Signalling. We will see that they are not, by constructing a counterexample. We will then propose a restriction on the family of composition rules to capture not only No Signalling but also {the orthogonality relations provided by Local Orthogonality} \cite{LO}. We show that the set of $\cQ_1$ correlations is invariant relative to the subset of composition rules compatible with this further restriction. 

One of the composition rules explored in this work is defined in terms of the abstract notion of a scenario's \textit{completion} (explained in~\ref{sec:completion}). Explicitly characterizing a scenario's completion is generally a difficult task. We conjecture here, however, that the completion-based composition protocol is equivalent to a newly-defined but easy-to-compute prodocut we call the disjunctive FR-product, building on the (now disproven) Conjecture C.2.6 of Ref. \cite{AFLS}. 

The paper is structured as follows. First we review the hypergraph formalism for contextuality scenarios \cite{AFLS}. In Section \ref{sec:compositionprereq} we discuss different notions of hypergraph compositions. {The full mathematical definitions of the various composition rules are, however, consigned to Appendix \ref{ap:multiprod}, in the interest of narrative flow.} In Section \ref{se:q1multi} we show that the set of $\cQ_1$ correlations in a given composite scenario depends on the particular choice of composition rule which was used. In Section \ref{se:diss} we discuss how to  {identify} those constructive composition rules which capture  {the orthogonality relations given by} Local Orthogonality, such that $\cQ_1$ models remain invariant under them. 

\section{Hypergraph Formalism}

In this section we review the basics of the hypergraph formalism for contextuality laid out in Ref. \cite{AFLS}. 

A contextuality scenario \cite{AFLS} is defined as a hypergraph $H = (V, E)$ whose vertices $v \in V$ correspond to the events in the scenario, i.e. the possible answers of the questions that can be asked to the system in the particular experimental setup. Each \textit{event} represents an outcome obtained from a device, {after it} received some input or ``measurement choice''. Every hyperedge\footnote{ {For simplicity we will refer to these hyperedges merely as edges whenever the context makes it clear.} } $e \in E$ in a non-composite contextuality scenario is a collection of events representing all the possible outcomes given a particular measurement choice. Note that both measurement outcomes and measurement choices can be vector-valued, as  {it} is naturally the case when describing the simultaneous measurement of different properties.

In the hypergraph formalism based on Ref. \cite{AFLS} we assume that every measurement set is complete, in the sense that if the measurement corresponding to $e$ is performed then exactly one of the outcomes corresponding to $v \in e$ is always obtained. Note that measurement sets may have non-trivial intersection: when an event appears in more than one hyperedge, this represents the idea that the two different operational outcomes should be thought of as equivalent.

A correlation -- hereafter \term{probabilistic model} -- on a contextuality scenario is an assignment of a probability-indicating number to each event, $p\,:\, V \, \to \, [0,1]$, hence denoting the relative frequency of the possible outcome of every measurement. This definition presupposes that the probability for a given event $v$ is independent of the measurement $e$ it is sampled from. 

Since the measurements are complete, every probabilistic model $p$ over the contextuality scenario $H$ satisfies the normalisation condition $\sum_{v \in e} p(v) = 1$ for every $e \in E$. The set of all possible probabilistic models on a contextuality scenario is denoted by $\mathcal{G}(H)$. In addition to imposing normalisation, the hyperedges also define the notion of orthogonality (a.k.a. exclusiveness) among events: $v$ and $w$ are orthogonal whenever there exists a hyperedge $e$ that contains both. Orthogonal events may be understood as contradictory counterfactual possibilities; this is elaborated in Section~\ref{sec:introduceLO}.

Several {stricter sets of} probabilistic models have been studied, depending on the nature of the system and, accordingly, the rule used to assign probabilities \cite{AFLS}. The definitions of classical and quantum probabilistic models are briefly reviewed in Appendix \ref{ap:models}, and we denote their corresponding sets of probabilistic models by $\cC(H)$ and $\cQ(H)$ respectively. Kochen and Specker \cite{KS} showed that there exist scenarios that admit quantum models but no classical models. More generally, however, there are many contextuality scenarios for which some models are inside the quantum set but outside the (nonempty) classical set, i.e. $\cC(H) \subsetneq \cQ(H)$ for some $H$. This general phenomenon is referred to as (quantum) \textit{contextuality}, in which the set of classical models are deemed the \textit{noncontextual models}.

In this work we {further focus} on the set of probabilistic models called $\cQ_1(H)$, which is particular relaxation of the quantum set.

\begin{defn}\label{q1mdef} $\cQ_1$ \textbf{models} \cite[Prop. 6.3.1]{AFLS} \\
Let $H$ be a contextuality scenario. An assignment of probabilities $p: V(H)\to [0,1]$ is a $\cQ_1$ model
if and only if there exists a Hilbert space $\cH$ upon which live some positive-semidefinite quantum state $\rho$ and positive-semidefinite projection operators $P_v$ associated to every $v\in V$ such that
\begin{equation}
\label{aqmeas}
1=\mathrm{tr}\left( \rho \right),\;\;\;\;p(v) = \mathrm{tr}\left( \rho P_v \right) \;\forall v\in V(H),\;\;\textrm{and}\;\;\;\;\sum_{v\in e} P_v \leq \mathbbm{1}_{\mathcal{H}} \;\forall e\in E(H).
\end{equation}
The set of all these models is denoted $\mathcal{Q}_1(H)$.
\end{defn}

The set of $\cQ_1$ models is physically relevant in several aspects. Firstly, it is closely related to the set of ``Almost-Quantum-Correlations" \cite{AQ}, which in Bell scenarios satisfies all the principles\footnote{With the potential exception of Information Causality.} proposed so far to try and single out quantum correlations. Indeed, the set of ``Almost-Quantum-Correlations" is \emph{defined} as the set of $\cQ_1$ models when applied to specific composite contextuality scenarios \cite{AFLS};  {we comment} on this connection further in next section. 
Valuably, $\cQ_1$ models can be effectively characterized by a semidefinite program \cite{AFLS}, hence striking a nice balance between closely approximating quantum correlations and being computationally tractable.

\section{Composition of Contextuality Scenarios}\label{sec:compositionprereq}

A natural question addressed in Ref. \cite{AFLS} is how to represent two independent systems as a single scenario, i.e. how two hypergraphs might combine into a joint one. A particular property that  {we demand} of this joint scenario is that the allowed general probabilistic models\footnote{The general models associated with contextuality scenarios are known, in nonlocality literature, as general probabilistic theory (GPT) correlations.} $\cG$ satisfy the so called \term{No Signalling} principle (NS): that the single-party\footnote{Here, ``party'' refers to each single component of the composite scenario.} marginals of every probabilistic model ought to be well-defined. To make this statement more precise, let us denote by $H_A$ the first scenario, with events labeled by $v_A$ and measurements by $e_A$, and by $H_B$ the second scenario, with events labeled by $v_B$ and measurements by $e_B$.
The events in this joint scenario are hence labeled by $(v_A,v_B)$. 
Then, the No Signalling principle states that if $p(v_A,v_B) \in \cG(H_{AB})$, where $H_{AB}$ is the joint scenario, then $\sum_{v_B \in e_B} p(v_A,v_B) $ should not depend on $e_B$, and similarly $\sum_{v_A \in e_A} p(v_A,v_B) $ should not depend on $e_A$. A composition rule which captures No Signalling, therefore, is a composition protocol which gives rise to hypergraphs innately implying the No Signalling principle on its general probabilistic models.

One can use elementary linear algebra to check that a candidate composite hypergraph correctly captures the No Signalling constraints. Conversely, one can check that the normalization of probability along a particular hyperedge in a composite hypergraph \emph{follows} from local normalization (that is, the normalization constraints in the scenarios being composed) and No Signalling. This is accomplished in practice by expressing hypergraphs as matrices with 0/1 entries, with rows corresponding to hyperedges and columns corresponding to vertices, as  {we elaborate} in the next few subsection.

\subsection{Linear Algebra of Hypergraphs}\label{sec:linearalgebra}
 
With a slight abuse of notation, we hereafter use $E(H)$ to denote both the matrix representation of $H$, and its set of hyperedges. 
 In matrix representation, the row corresponding to a particular hyperedge is such that a ``1" is assigned to a column if and only if the vertex corresponding to that column belongs to that particular hyperedge. An example of a hypergraph in matrix notation can be found in Appendix~\ref{ap:q1comp}.

Firstly, any composite scenario $H_{\textrm{Composite}}=\otimes_i H_i$, no matter the details of the composition protocol,  {has} a vertex set given by $V\left(H_{\textrm{composite}}\right) \defin \prod_i V(H_i)$,  {where $\prod_i$ hereon denotes cartesian product}. Second, a model $p$ is allowed in the scenario $H$ only if $E(H) \cdot \vec{p} = \vec{\bf{1}}$, where {$\vec{p} = \{p(v_k)\}_k$} and $\vec{\bf{1}}$ is a column-vector of all-ones. 

\clearpage Therefore, a hypergraph correctly\footnote{{By `correctly' we mean that the hyperedges in the composite hypergraph impose \textit{all and only} the restrictions on general probabilistic models captured from normalization and No Signalling.}} captures normalization and No Signalling if and only if the reduced row echelon form of $M_{\textrm{Normalization}}^{\vec{\bf{1}}} \cup M_{\textrm{NoSignalling}}^{\vec{\bf{0}}}$ --- both defined below --- is the same as the reduced row echelon form of the hypergraph's matrix representation supplemented by a column of ones. Here, the $\cup$ operation denotes taking the union of the rows of the matrices, and $M^{\vec{\bf{0}}}, M^{\vec{\bf{1}}}$ denote padding a matrix with a column of all-zeroes or all-ones respectively. As  {we'll see} below, the matrices $M_{\textrm{Normalization}}$ and $M_{\textrm{NoSignalling}}$ have the same column-indices as $E(H_{\textrm{Composite}})$, namely the vertices of $H_{\textrm{Composite}}$.
 {Hence, $\otimes_{i} H_i$ formally} captures No Signalling if and only if
\begin{align}
\mathtt{rref}\!\left(E^{\vec{\bf{1}}}(\otimes_i H_i)\right)=\operatorname{\mathtt{rref}}\!\left(M_{\textrm{Normalization}}^{\vec{\bf{1}}} \cup M_{\textrm{NoSignalling}}^{\vec{\bf{0}}}\right).
\end{align}

\begin{sloppypar}
The \term{composite normalization matrix} $M_{\textrm{Normalization}}$ is defined as follows. A row in $M_{\textrm{Normalization}}$ consists of 0/1 entries, such that the rows in $M_{\textrm{Normalization}}$ are effectively hyperedges, which we call the \emph{normalization hyperedges}. $M_{\textrm{Normalization}}$ is designed to capture normalization of composite vertex sets based on normalization of the local vertex sets. A normalization hyperedge  {hence corresponds to the vertex set} $\Pi_{i} e_{k_i}$,  for some fixed set of local hyperedges $e_{k_i}\in E_i\;\forall_i$. The full normalization matrix is constructed by considering every possible tuple $\{e_{k_i}\}_i$ of local hyperedges.
\end{sloppypar}

The \term{composite No Signalling matrix} $M_{\textrm{NoSignalling}}$ is defined as follows. A row in $M_{\textrm{NoSignalling}}$ consists of -1/0/+1 entries. Each row in $M_{\textrm{NoSignalling}}$ captures a No Signalling equality, such that $M_{\textrm{NoSignalling}} \cdot \vec{p} = \vec{\bf{0}}$,  {as explained in the following}. Given a composite scenario of $n$ state spaces $H_j$, $j=1\ldots n$, a No Signalling equality is  {equivalently} parameterized by
\begin{samepage}\begin{compactitem}
\item[(i)] the choice of a local scenario $H_k$,
\item[(ii)] the choice of some pair of hyperedges in that scenario, $e^1_k, e^{2}_k \in E(H_k)$,
\item[(iii)] the choice of some fixed local vertices $v_j \in V(H_{j\neq k})$, one for each of the local scenarios other than $H_k$. 
\end{compactitem}\end{samepage}

Given (i) -- (iii), a No Signalling equality reads
\begin{align}\label{eq:normmat}
\sum_{v_k \in e^1_k} p(v_{1},..., v_{n}) = \sum_{v_k \in e^2_k} p(v_{1},..., v_{n}).
\end{align}
From Eq.~\eqref{eq:normmat}, such a No Signalling equality translates into a row in $M_{\textrm{NoSignalling}}$ as follows: We assign a $+1$ to those columns labelled by events ${(v_{1},..., v_{n})}$ where ${v_k \in e^1_k \setminus e^2_k}$, we assign $-1$ to those columns where ${v_k \in e^2_k \setminus e^1_k}$, and we assign $0$ to all the remaining columns.

\subsection{Equivalence of Hypergraphs under Completion}\label{sec:completion}
The reduced echelon form of a matrix representation of a hypergraph becomes useful when discussing the notion of a scenario's \term{completion}. The completion of a contextuality scenario is {the hypergraph} constructed by starting with {that of} the original scenario and then adding as hyperedges every set of vertices $e_V$ having the property that $\sum_{v \in e_V} p(v) = 1$ for all $p \in \mathcal{G}(H)$. These additional hyperedges $e_V$ present in the completion but absent in the original scenario are called \textit{virtual} hyperedges.  

\begin{figure}[t]
\begin{center}
\begin{tikzpicture}
\node[draw,shape=circle,fill,scale=.5] (a) at (0:2cm) {} ;
\node at (72:2.6cm) {$u_1$};
\node at (288:2.6cm) {$u_2$};
\node[draw,shape=circle,fill,scale=.5] (b) at (72:2cm) {} ;
\node[draw,shape=circle,fill,scale=.5] (c) at (144:2cm) {} ;
\node[draw,shape=circle,fill,scale=.5] (d) at (216:2cm) {} ;
\node[draw,shape=circle,fill,scale=.5] (e) at (288:2cm) {} ;
\draw[thick,blue,rotate=36] (0:1.63cm) ellipse (.4cm and 1.6cm) ;
\draw[thick,blue,rotate=108] (0:1.63cm) ellipse (.4cm and 1.6cm) ;
\draw[thick,blue,rotate=180] (0:1.63cm) ellipse (.4cm and 1.6cm) ;
\draw[thick,blue,rotate=252] (0:1.63cm) ellipse (.4cm and 1.6cm) ;
\draw[thick,blue,rotate=324] (0:1.63cm) ellipse (.4cm and 1.6cm) ;
\draw[thick,orange,dashed] (.6cm,0) ellipse (.6cm and 2.2cm) ;
\end{tikzpicture}
\end{center}
\caption{Example of an incomplete contextuality scenario, i.e. a hypergraph with virtual hyperedges (dashed) Only one virtual hyperedge is highlighted, but this scenario has five virtual hyperedges. Novel orthogonality relations are induced when adding virtual hyperedges to this scenario, such as $u_1\bot u_2$.}
\label{pentagonvirt}
\end{figure}
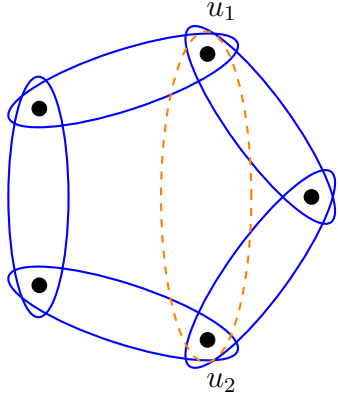

A candidate hyperedge $e$ is a virtual hyperedge of scenario $H$ if and only if \begin{align}\label{eq:virtualedgedef}
e\not\in E(H)\:\:\text{ and }\:\:\mathtt{rank}\!\left(E^{\vec{\bf{1}}}(H)\cup{e^{\vec{\bf{1}}}}\right)=\mathtt{rank}\!\left(E^{\vec{\bf{1}}}(H)\right).
\end{align}
An example of virtual hyperedges in matrix notation can be found in Appendix~\ref{ap:q1comp}. 
Now, it is possible for two distinct hypergraphs to nevertheless have coinciding reduced row echelon forms of their matrix representation; this occurs if and only if the hypergraphs are equivalent up to virtual hyperedges. 
In this case, the hypergraphs imply precisely the same set of general probabilistic models, and hence we say that they are equivalent under \textbf{completion}:
\begin{align*}
H \text{ is equivalent under completion to } H' \text{ iff } \mathtt{rref}\!\left(E^{\vec{\bf{1}}}(H)\right)=\mathtt{rref}\!\left(E^{\vec{\bf{1}}}(H')\right).
\end{align*} 

The completion of a hypergraph, which we denote with overline, is the operation of adding all its possible virtual hyperedges. Equivalently, the completion may be characterized in terms of reduced row-echelon form of $E(H)$: it is the largest collection of hyperedges - i.e. rows made up only of zeroes and ones in matrix notation - such that the reduced row echelon is preserved, i.e. no further hyperedges can be added to $\overbar{H}$ without increasing the rank of $E(\overbar{H})$. Formally, 
\begin{align}\label{defcompletion}\begin{split}
\overbar{H}&\!\defin  \text{ the unique hypergraph such that }\mathtt{rref}\!\left(E^{\vec{\bf{1}}}(\overbar{H})\right)=\mathtt{rref}\!\left(E^{\vec{\bf{1}}}(H)\right)
\\&\text{and }\; \nexists_{e\notin E(\overbar{H})} \text{ such that }\:\:\mathtt{rank}\!\left(E^{\vec{\bf{1}}}(\overbar{H})\cup{e^{\vec{\bf{1}}}}\right)=\mathtt{rank}\!\left(E^{\vec{\bf{1}}}(\overbar{H})\right).
\end{split}\end{align}
Since we treat every hyperedge $e$ as a row vector, one should therefore understand $e^{\vec{\bf{1}}}$ appearing in Eq.~\eqref{defcompletion} as simply denoting appending a single ``1" onto the end of vector $e$.

Checking if a hypergraph is self-complete is a closed but difficult computational task\footnote{An inefficient approach to completing a hypergraph is to generate \emph{all} hyperedges not in the hypergraph, and then to asses if they are virtual hyperedges of the starting hypegraph, pursuant to Eq.~\eqref{eq:virtualedgedef}.}, except for certain special cases. For instance, if a hypegraph has only non-intersecting hyperedges then it a a-priori self-complete. The authors are not aware of any reasonable computational algorithm for constructing the completion of a generic incomplete hypergraph. In this work we will generally refer to $\overbar{H}$ as an abstract idea, although we will show that the completion of composite hypegraph can be explicitly constructed given certain promises about the individual hypegraph. Note that a sufficient criterion for $H_1$ and $H_2$ to have the same completion is $E(H_1) \subseteq E(H_2)$ together with ${\mathtt{rank}\!\left(E^{\vec{\bf{1}}}(H_1)\right)=\mathtt{rank}\!\left(E^{\vec{\bf{1}}}(H_2)\right)}$. 

\subsection{The Orthogonality of Event Pairs in a Contextuality Scenario}\label{sec:introduceLO}

Preliminary to an assessment of composition rules we first establish some terminology regarding orthogonality relationship between events in a contextuality scenario. A pair of events are mutually exclusive in a given contextuality scenario if both events are common to any single hyperedge. If some hyperedge in $H$ contains both $v$ and $w$ then we say that $v$ and $w$ are \term{orthogonal} in $H$, which we denote $H\mathpunct{\Rightarrow}v\bot w$  {or $v\bot_H w$}. Formally,
\begin{align}\label{eq:orthodef}
H\mathpunct{\Rightarrow}v\bot w\quad\text{ iff }\quad\exists\: e\!\in\! E(H)\:\:\text{ such that }\:\: \{v,w\}\mathpunct{\subset} e\,.
\end{align}
In single-partite scenarios, events common to some hyperedge represent alternative possible outcomes of a single measurement. This generalize to composite scenarios as well, wherein orthogonality indicates the exclusivity of possibility: if $v\bot_H w$ then $v$ and $w$ are counterfactually contradictory.

\begin{figure}
\begin{center}
\begin{tikzpicture}
\node[draw,shape=circle,fill,scale=.5] at (0,0) {} ;
\node[draw,shape=circle,fill,scale=.5] at (1.4,0) {} ;
\node[draw,shape=circle,fill,scale=.5] at (0,1.5) {} ;
\node[draw,shape=circle,fill,scale=.5] at (1.4,1.5) {} ;
\draw[thick,blue] (0.7,0) ellipse (1.4cm and .3cm) ;
\draw[thick,blue] (0.7,1.5) ellipse (1.4cm and .3cm) ;
\end{tikzpicture}
\end{center}
\caption{Hypegraph representing the contextuality scenario where a device measures on the system in one of two ways, each of which gives dichotomic outcomes. Later on this scenario is understood as the single-party contextuality scenario $B_{1,2,2}$.}
\label{the122example}
\end{figure}
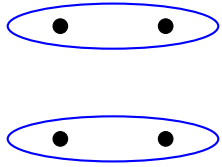

As an example, imagine that an experimenter Alice chooses to perform measurements on some system\footnote{The measurements can be disturbative, such that the pristine system is contaminated by the action of a measurement. Disturbative measurements prevent asking too many different questions about the system; rather, the experimenter must choose one among multiple possible measurements  to perform.} using a measurement apparatus which could be configured in one of two ways (see Fig.~\ref{the122example}). Let the events $(0|0)$ and $(1|0)$ respectively indicate getting her seeing outcome zero versus obtaining outcome one from her measurement apparatus \emph{when the measurement apparatus is configured in the default setting}. Let the events $(0|1)$ and $(1|1)$ again indicate outcomes zero and one respectively, but with the specification that Alice was using the alternative setting configuration. Then, $(0|0)\bot(1|0)$ because if Alice gets outcome zero using the default setting then she \emph{could not} have gotten outcome one, and vice versa. This is the sense in which orthogonal events are contradictory counterfactuals. On the other hand, if Alice gets outcome zero using the default setting it does not necessarily mean that she wouldn't have obtained outcome one if she had selected the alternative setting, and hence $(0|0)\not\bot(1|1)$.

As orthogonality is a strictly pairwise relationship, we can represent the orthogonality relations implied by a contextuality scenario as a simple undirected graph, which we call the \term{orthogonality graph} of $H$, denoted $\OG(H)$. One can construct $\OG(H)$ from $E(H)$, but not vice versa. Formally, every node in $H$ is a node in $\OG(H)$, and orthogonality pursuant to $H$ defines the adjacency condition $\OG(H)$:
\begin{align}\label{eq:orthographdef}
v \mathpunct{\sim} w \:\text{ is an edge of } \OG(H)\quad\text{ iff }\quad H\mathpunct{\Rightarrow}v\bot w\,.
\end{align}
Note that different contextuality scenarios, i.e. different hypergraphs, can admit identical sets of orthogonality relations, and hence can share a common orthogonality graph. In the formalism of Ref.~\cite{AFLS} the preferred nomenclature is to speak of non-orthogonality graphs, such that $v\mathpunct{\sim} w$ is in $\NOG(H)$ if and only if $H\mathpunct{\Rightarrow}v\not\bot w$. These two graphical representations of orthogonality are readily interconvertible, as they are simply graphical complements.

\begin{figure}[t]
\begin{center}
\subfigure[Contextuality scenario $H$.]{
\begin{tikzpicture}[scale=1.5]
\draw[thick,blue] plot [smooth cycle,tension=.8] coordinates { (0,1.9) (-0.65,0.75) (0.65,0.75) } ;
\draw[thick,blue] plot [smooth cycle,tension=.8] coordinates { (-0.5,1.05) (-1.15,-0.1) (0.15,-0.1) } ;
\draw[thick,blue] plot [smooth cycle,tension=.8] coordinates { (0.5,1.05) (-0.15,-0.1) (1.15,-0.1) } ;
\draw[thick,blue] (0,0.59) circle (1.05cm);
\draw[thick,blue] (0,0.59) circle (1.25cm);
\draw[thick,dashed,orange,rotate around={30:(0.5,0.866)}] (0.5,0.866) ellipse (.2cm and 1.6cm) ;
\draw[thick,dashed,orange,rotate around={-30:(-0.5,0.866)}] (-0.5,0.866) ellipse (.2cm and 1.6cm) ;
\draw[thick,dashed,orange,rotate around={90:(0,0)}] (0,0) ellipse (.2cm and 1.6cm) ;
\draw[thick,dashed,orange] plot [smooth cycle,tension=.8] coordinates { (0,-0.15) (-0.63,0.95) (0.63,0.95) } ;
\node[draw,shape=circle,fill,scale=.5,label=below:{$v_1$}] at (0,1.7321) {};
\node[draw,shape=circle,fill,scale=.5,label=above left:{$v_2$}] at (-0.5,0.866) {};
\node[draw,shape=circle,fill,scale=.5,label=above right:{$v_3$}] at (0.5,0.866) {};
\node[draw,shape=circle,fill,scale=.5,label=above right:{$v_4$}] at (-1,0) {};
\node[draw,shape=circle,fill,scale=.5,label=above:{$v_5$}] at (0,0) {};
\node[draw,shape=circle,fill,scale=.5,label=above left:{$v_6$}] at (1,0) {};
\end{tikzpicture}}
\hskip 0.5cm
\subfigure[ $\OG(H) = \OG(\overline{H})$.]{
\begin{tikzpicture}[scale=1.5]
\node[draw,shape=circle,fill,scale=.5,label=below:{$v_1$}] at (0,1.7321) {};
\node[draw,shape=circle,fill,scale=.5,label=above left:{$v_2$}] at (-0.5,0.866) {};
\node[draw,shape=circle,fill,scale=.5,label=above right:{$v_3$}] at (0.5,0.866) {};
\node[draw,shape=circle,fill,scale=.5,label=above right:{$v_4$}] at (-1,0) {};
\node[draw,shape=circle,fill,scale=.5,label=above:{$v_5$}] at (0,0) {};
\node[draw,shape=circle,fill,scale=.5,label=above left:{$v_6$}] at (1,0) {};
\draw (0,1.7321)  -- (-0.5,0.866) --  (-1,0)  -- (0,0) -- (1,0)  --  (0.5,0.866) --  (-0.5,0.866) -- (0,0) --  (0.5,0.866) -- cycle;
\draw (0,0.59) circle (1.15cm);
\node at (2,0) {};
\node at (-2,0) {};
\end{tikzpicture}}
\end{center}
\caption{(a) A contextuality scenario $H$ that is incomplete yet ortho-stable. It has 6 vertices, 4 hyperedges (solid blue lines), and 4 virtual hyperedges (dashed orange lines). (b) Orthogonality graph of both $H$ and its completion $\overline{H}$.}
\label{fig:orthoest}
\end{figure}
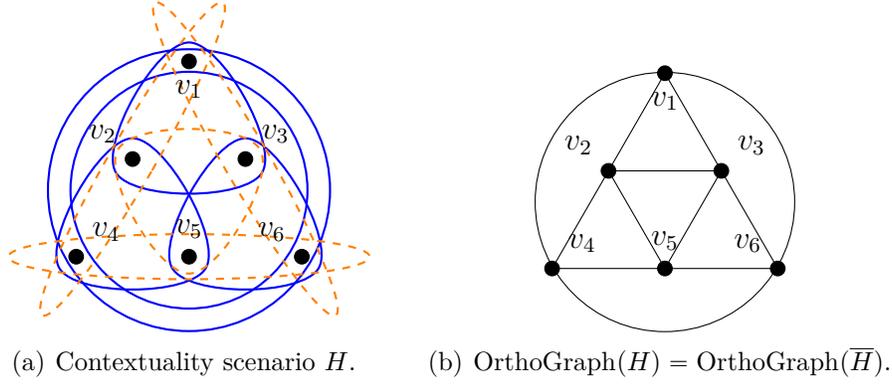

One can conceive of a contextuality scenario $H$ such that two events are not orthogonal according to $H$ yet the two events \emph{are} orthogonal in the scenario's completion, i.e. $H\mathpunct{\Rightarrow}v\not\bot w$ while $\overbar{H}\mathpunct{\Rightarrow}v\bot w$. This occurs when $v$ and $w$ do not appear together in a single $e\!\in\!E(H)$ but $v$ and $w$ are co-members of some virtual hyperedge of $H$. Such hypegraphs have orthogonality relations which change under completion; an example is illustrated in Fig.~\ref{pentagonvirt}. These contextuality scenarios  {might be} awkward to interpret physically; we therefore introduce notation to classify hypegraphs according to satisfaction (or lack thereof) of this property. We say that a hypegraph is \term{ortho-stable} whenever ${\OG(H)=\OG(\overbar{H}})$, and ortho-unstable otherwise. 
 { Fig.~\ref{fig:orthoest} presents an example of an incomplete yet ortho-stable contextuality scenario. There, $H$ consists of 6 vertices $V(E)=\{v_i\}_{i=1\ldots6}$ and 4 edges:
\begin{align*}
\begin{tabular}{c | c c c c c c}
 & $v_1$ & $v_2$ & $v_3$ & $v_4$ & $v_5$ & $v_6$ \\
 \hline
 \(\! e_1 \!\) & 1 & 1 & 1 & \({ \scriptscriptstyle ^0 }\) & \({ \scriptscriptstyle ^0 }\) & \({ \scriptscriptstyle ^0 }\) \\
 \(\! e_2 \!\) & \({ \scriptscriptstyle ^0 }\) & 1 &  \({ \scriptscriptstyle ^0 }\) & 1 & 1 & \({ \scriptscriptstyle ^0 }\) \\
 \(\! e_3 \!\) & \({ \scriptscriptstyle ^0 }\) & \({ \scriptscriptstyle ^0 }\) & 1 &  \({ \scriptscriptstyle ^0 }\) & 1 & 1 \\
 \(\! e_4 \!\) & 1 & \({ \scriptscriptstyle ^0 }\) & \({ \scriptscriptstyle ^0 }\) & 1 &  \({ \scriptscriptstyle ^0 }\) & 1 
\end{tabular}
\end{align*}
It is a straightforward calculation to check that the list of virtual hyperedges in $H$ has four elements: 
\begin{align*}
\begin{tabular}{c | c c c c c c}
 & $v_1$ & $v_2$ & $v_3$ & $v_4$ & $v_5$ & $v_6$ \\
 \hline
 \(\! e_5 \!\) & \({ \scriptscriptstyle ^0 }\) & \({ \scriptscriptstyle ^0 }\) & \({ \scriptscriptstyle ^0 }\) & 1 & 1 & 1 \\
 \(\! e_6 \!\) & \({ \scriptscriptstyle ^0 }\) & 1  & 1 &  \({ \scriptscriptstyle ^0 }\) & 1 & \({ \scriptscriptstyle ^0 }\) \\
 \(\! e_7 \!\) & 1 &  \({ \scriptscriptstyle ^0 }\) & 1 & \({ \scriptscriptstyle ^0 }\) & \({ \scriptscriptstyle ^0 }\)  & 1 \\
 \(\! e_8 \!\) & 1 & 1 & \({ \scriptscriptstyle ^0 }\) &  1 &\({ \scriptscriptstyle ^0 }\) &  \({ \scriptscriptstyle ^0 }\)
\end{tabular}
\end{align*}
The orthogonality relations given by these virtual hyperedges, however, are already imposed by $E(H)$. Hence, $\overbar{H}$ has the same orthogonality graph as $H$. }

\bigskip

When one composes multiple hypegraphs to form a single composite contextuality scenario, each event in the composite scenario is effectively a collection of local events; i.e. a global (a.k.a composite) event is identically the simultaneous occurrence of an event in each of the individual (a.k.a local) scenarios. We here represent composite events as a stack of local events: for example, the occurrence of the tripartite Bell scenario event $\biggl(\!\begin{smallmatrix}1|0\\0|1\\1|1\end{smallmatrix}\!\biggr)$ should be read as ``Alice's got outcome 1 using setting 0, Bob got outcome 0 using setting 1, and Charlie's got outcome 1 using setting 1." 

There is a natural notion of orthogonality for such multipartite events, namely \term{Local Orthogonality} (\term{LO})\footnote{ {For simplicity, hereafter we denote by LO the orthogonality relations implied by the Local Orthogonality principle rather than the whole principle itself.}}. In any composite scenario generated by the individual hypegraphs $H_1,...H_n$, the $n$-partite composite events $v\mathpunct{\defin}\biggl(\!\begin{smallmatrix}v_1\vspace{-1ex}\\\vdots\\v_n\end{smallmatrix}\!\biggr)$ and $w\mathpunct{\defin}\biggl(\!\begin{smallmatrix}w_1\vspace{-1ex}\\\vdots\\w_n\end{smallmatrix}\!\biggr)$ are locally orthogonal \cite{LO} if and only the composite events  {involve} some corresponding pair of local events $\{v_k, w_k\}$ which are orthogonal pursuant to their relevant individual hypegraph, i.e. 
\begin{align}\label{eq:newdefLO}
H_1\otimes ...\otimes H_n\mathpunct{\Rightarrow}v\underaccent{\mathrm{LO}}{\bot}w \quad\text{ iff }\quad \exists\; k\:\:\text{ such that }\:\: H_k\mathpunct{\Rightarrow}v_k\bot w_k\,.
\end{align} 
The LO notion of orthogonality, while intuitive, is independent of the standard definition. The composite events $v$ and $w$ are only \emph{conventionally} orthogonal if they appear together in some hyperedge in the composite scenario, and the definition of the hyperedges in the composite scenario depends on the choice of product rule used for composition! We therefore introduced the new notation 
$v\underaccent{\mathrm{LO}}{\bot}w$
 for indicating that $w$ is \emph{locally} orthogonal to $v$, in contrast to the usual $v\bot w$ for indicating conventional orthogonality.

If the set of Local Orthogonality relations between composite events is included among the conventional orthogonality relations, then we say that the composition \term{captures LO}. If the conventional relations match up perfectly with LO relations, i.e.~Local Orthogonality is captured and  {in addition} the conventional definition of orthogonality does not imply any orthogonality beyond LO, then we say that the composition \term{faithfully captures LO}. Different methods of composition, even when yielding different final composite hypergraphs, can induce a single common orthogonality graph. In particular, all compositions which faithfully capture LO share a special orthogonality graph, defined as
\begin{align}\begin{split}\label{eq:minLOgraph}
\operatorname{FaithfulOrthoGraph}(H_1,...,H_n) =  
\OG(H_1)\Asterisk...\Asterisk\OG(H_n)
\end{split}\end{align}
where $\Asterisk$ indicates the \term{disjunctive graphical product}, also known the co-normal graphical product\footnote{Readers with an information-theoretic background might be familiar with the \term{strong graphical product} $\boxtimes$, which is used to define a graph's Shannon capacity. Formally, $g_1 \Asterisk g_2 = !((!g_1)\boxtimes(!g_2))$, where $!g$ denotes the graphical compliment in which adjacent vertices are made disconnected and edges are added between all pairs of disconnected vertices.}. The definition of the disjunctive product of graphs parallels that of LO: 
${\biggl(\!\begin{smallmatrix}v_1\vspace{-1ex}\\\vdots\\v_n\end{smallmatrix}\!\biggr) \sim \biggl(\!\begin{smallmatrix}w_1\vspace{-1ex}\\\vdots\\w_n\end{smallmatrix}\!\biggr)}$ is an edge of $g_1\Asterisk...\Asterisk g_n$ if and only if there exists a $k$ such that $v_k\mathpunct{\sim} w_k$ is an edge of $g_k$. The disjunctive graphical products is associative, in that $g_1 \Asterisk g_2 \Asterisk g_3 = (g_1 \Asterisk g_2)\Asterisk g_3 = g_1 \Asterisk (g_2 \Asterisk g_3)$. In summary, a composition faithfully captures LO if and only if the composition's orthogonality graph coincides with the faithful orthogonality graph define above.

In the following sections we will filter different candidate composition rules into those protocols which faithfully capture LO and those which do not. We will show that $\cQ_1$ models are sensitive to the choice of product rule only when the choice leads to different orthogonality relations among compositive events, such that $\cQ_1$ models are invariant relative to the subset of composition rules which faithfully capture LO.

\subsection{A Hierarchy of Composition Rules}\label{sec:manyrules}

How to explictly construct a joint scenario $H_{AB}$ from two individual ones $H_A$ and $H_B$ in such a manner as to correctly capture No Signalling and normalization was elegantly solved by Foulis and Randall \cite{FR}, via their Foulis-Randall product: $H_{AB} = H_A {}^{\FR}\otimes H_B$. Although we will consider of a variety of distinct protocols for composing multiple $(n>2)$ individual ones, it is important to note that almost\footnote{The bipartite FR-product is not equal to its own completion if any of the local hypergraphs are themselves incomplete, i.e. $\overbar{H_A{^{\FR}\otimes} H_B}\neq H_A{^{\FR}\otimes} H_B \text{ if }\overbar{H_A}\neq H_A \text{ or }\overbar{H_B}\neq H_B$. We conjecture, however, that $\overbar{H_A{^{\FR}\otimes} H_B}= H_A{^{\FR}\otimes} H_B \text{ if }\overbar{H_A}= H_A \text{ and }\overbar{H_B}= H_B$, such as is locally the case for Bell scenarios.} the entire plethora of multipartite composition methods reduce to precisely the same protocol when considering $(n=2)$, namely the unambiguous bipartite FR-product, whose definition is presented in Appendix \ref{ap:prod}. 

The result by Foulis and Randall, however, is not the ultimate answer to the problem, because the FR-product is not associative when considering more than two contextuality scenarios. Different associative extensions of ${^{\FR}\otimes}$ were explored in Ref. \cite[Appendix C]{AFLS}. In this work  {we comment} primarily on five (associative) hypergraph composition protocols:\newline
\begin{tabular}{lllr}
1. & the \textit{minimal} FR-product,& denoted & $\quad^{\min}\otimes$,\\
2. & the \textit{common} FR-product, & denoted & $\quad^{\comm}\otimes$,\\
3. & the \textit{maximal} FR-product, & denoted & $\quad^{\max}\otimes$,\\
4. & the \textit{disjunctive} FR-product, & denoted & $\quad^{\disj}\otimes$,\\
5. & and the \textit{completed} FR-product, & denoted & $\quad\overbar{\otimes}$.
\end{tabular}\newline
In the interest of {focusing} on new results, and since most of the five protocols have already been introduced in Ref. \cite{AFLS}, we relegate all definitions of the individual protocols to Appendix~\ref{ap:multiprod}. 

Each choice of product in the list gives rise to its own form of composite hypergraphs. Importantly, the five protocols form a well-ordered hierarchy of equivalent-under-completion hypergraphs, per Prop.~\ref{prop:hierarchyofedges} in Appendix~\ref{sec:proofs}. Formally,
\begin{align}\label{eq:hierarchy}
E(^{\min}\!\otimes_i H_i) \subseteq
E(^{\comm}\!\otimes_i H_i)\subseteq
E(^{\max}\!\otimes_i H_i)\subseteq 
E(^{\disj}\!\otimes_i H_i) \subseteq
E(\overbar{\otimes} H_i).
\end{align}
 {At} the same time
\begin{align}\label{eq:samecompletion}
\mathtt{rank}(E(\overbar{\otimes} H_i))=\mathtt{rank}(E(^{\min}\!\otimes_i H_i)),
\end{align}
consistent with the equivalence of all five composition protocols under completion.

The representation of Bell scenarios as contextuality scenarios also depends on the choice of composition rule, although no ambiguity arises until the Bell scenario consists of at least three parties.  {We denote} the hypergraph representation of a Bell scenario by $\cB_{n,m,d}^{\mathrm{prod}} = (V,E)$, where $n$ is the number of parties, $m$ the number of measurements they choose from, $d$ the number of outcomes  {each of those measurements has,} and $prod$ the choice of hypergraph product. As far as general models, classical models, and even quantum models are concerned, the choice or product is totally irrelevant, as proven in  Ref. \cite[Thm. C.2.4]{AFLS} and elaborated in Prop.~\ref{eq:modelsinvariant} of Appendix~\ref{sec:proofs}. That is,
\begin{align}
&\cG(\cB_{n,m,d}^{\min})=\cG(\cB_{n,m,d}^{\comm})=\cG(\cB_{n,m,d}^{\max})=\cG(\cB_{n,m,d}^{\disj})=\cG(\overbar{\cB_{n,m,d}})\label{eq:GPTequivalence}\\
&\cC(\cB_{n,m,d}^{\min})=\cC(\cB_{n,m,d}^{\comm})=\cC(\cB_{n,m,d}^{\max})=\cC(\cB_{n,m,d}^{\disj})=\cC(\overbar{\cB_{n,m,d}})\label{eq:classicalequivalence}\\
&\cQ(\cB_{n,m,d}^{\min})=\cQ(\cB_{n,m,d}^{\comm})=\cQ(\cB_{n,m,d}^{\max})=\cQ(\cB_{n,m,d}^{\disj})=\cQ(\overbar{\cB_{n,m,d}}).\label{eq:quantumequivalence}
\end{align}
For such models, then, all choices of product give rise to \emph{observationally} equivalent scenarios.

Invariance under choice of product cannot be expected for all models, however. In particular,  {in the next section we show} that $\cQ_1$ models are observationally distinct between the minimal FR-product and the common FR-product. Indeed, we explicitly confirm that
\begin{align}
\cQ_1(\cB_{3,2,2}^{\min})\neq\cQ_1(\cB_{3,2,2}^{\comm}).
\end{align}
On the other hand,  {we prove} that $\cQ_1$ models are invariant relative to the choice of $^{\comm}\!\otimes$, $^{\textrm{max}}\otimes$, or $^{\disj}\!\otimes$, i.e.
\begin{align}
\cQ_1(\cB_{n,m,d}^{\comm})=\cQ_1(\cB_{n,m,d}^{\max})=\cQ_1(\cB_{n,m,d}^{\disj}),
\end{align}
such that the forthcoming results in Sec. \ref{se:q1multi} may be summarized as finding
\begin{align}
\cQ_1(\cB_{3,2,2}^{\min})\supsetneq\cQ_1(\cB_{3,2,2}^{\comm})=\cQ_1(\cB_{3,2,2}^{\max})=\cQ_1(\cB_{3,2,2}^{\disj})\supseteq\cQ_1(\overbar{\cB_{3,2,2}}).
\end{align}

An important motivation to study $\cQ_1$ models is that the set of $\cQ_1$ models coincides with that of the ``Almost-Quantum-Correlations" \cite{AQ} for the joint scenario generated by $^{\comm}\!\otimes$, $^{\max}\!\otimes$, or $^{\disj}\!\otimes$ \cite[Sec. 6.4]{AFLS}. In Sec. \ref{se:q1multi} we will see that $\cQ_1$ does depend on the choice of product, although in Sec.~\ref{se:diss} we will argue that restricting  { tocomposition rules which capture LO} renders that subset observationally equivalent as far as $\cQ_1$ models are concerned.

\section{$\cQ_1$ Models in Multipartite Composite Systems}\label{se:q1multi}

In this section we show that the set of $\cQ_1$ models may depend on the choice of multipartite product, i.e.~the choice of composition protocol. 
The bipartite case is mostly (but not entirely) trivial, because in that case the composite hypergraphs pursuant to any protocol $^{\min}\!\otimes$ through $^{\disj}\!\otimes$ all coincide. Obviously, if two product choices both yield the same exact scenario, they will also yield the same set of $\cQ_1$ models, trivially. We therefore begin by exploring the next simplest case, namely three parties. We hereafter consider those composite Bell scenarios where each of the three parties has two disjoint dichotomic measurements, i.e. Bell scenarios which -- while composition-rule-dependent -- are all of the family $\cB_{3,2,2}^{\textrm{prod}}$.

\subsection{The Different Compositions of the 3 Parties}\label{sec:tripartiteBell}

Firstly note that, regardless of the choice of product, $\cB_{3,2,2}^{\textrm{prod}}$ has 64 vertices, which correspond to the eight different possible joint-measurement outcomes for each of the eight different possible (fixed) multipartite measurement contexts. The different product choices nevertheless lead to distinct hypergraphs:
\begin{compactitem}
\item $\cB_{3,2,2}^{\min}$ has \textbf{176} hyperedges,
\item $\cB_{3,2,2}^{\comm}$ has \textbf{488}, 
\item $\cB_{3,2,2}^{\max}$ has \textbf{680}, 
\item $\cB_{3,2,2}^{\disj}$ has \textbf{744}.
\end{compactitem}
As an example, let us discuss how to see that there are indeed 176 hyperedges in $\cB_{3,2,2}^{\min}$. First notice that there are two different kinds of measurements in the scenario: (i) those where all three measurement choices are fixed throughout the hyperedge, and (ii) those where the choice of measurement for one party depends nontrivially on the outcomes of the other two. Edges of type (i) are those of `cartesian product form', and hence there are $2^3=8$ of them. Edges of type (ii) consist of a choice of two parties that measure first (there are 3 such choices), a choice of their pair of measurements ($2^2=4$ such choices), and finally a choice of a function $f$ from their $4$ outcomes to a choice of measurement for the remaining party. Note that this function $f$ must be nontrivial  {(i.e.~it should not be a constant)}, since otherwise we would recover a hyperedge of type (i). There are hence $2^4-2 = 14$ such functions. The total number of hyperedges in $\cB_{3,2,2}^{\min}$ is then $8+3*4*14 = 176$. 

Since we always have a hierarchy of hypergraphs per Eq. (\ref{eq:hierarchy}), we also always have a hierarchy of allowed probabilistic models. We begin by a-priori recognizing that surely
\begin{align}
\cQ_1(\cB_{3,2,2}^{\min})\supseteq\cQ_1(\cB_{3,2,2}^{\comm})\supseteq\cQ_1(\cB_{3,2,2}^{\max})\supseteq\cQ_1(\cB_{3,2,2}^{\disj})\supseteq\cQ_1(\overbar{\cB_{3,2,2}}).
\end{align}

\begin{sloppypar}
\subsection{Observational Equivalence of $\cQ_1$ Models from Common to Disjunctive FR-Products}\label{sec:equivalenceabovecommon}
\end{sloppypar}

Before showing dependence of $\cQ_1$ on composition rule, however, we pause to show that the choices of composition methods between (and including) the common product and the disjunctive product are \emph{observationally equivalent}  {regarding} the set of $\cQ_1$ models. That is,
\begin{align}
{\cQ_1(^{\comm}\!\otimes_i H_i)=\cQ_1(^{\disj}\!\otimes_i H_i)}
\end{align}
holds in general, even for arbitrary contextuality scenarios, i.e. the invariance of  $\cQ_1$ models under such products is not limited to Bell scenario. The proof is discussed below.

\begin{sloppypar}
Per Ref. \cite[Prop.~6.3.2]{AFLS}, a probabilistic model $p\in\mathcal{G}(H)$ is in $\mathcal{Q}_1$ if and only if $\vartheta(\NOG(H),p)=1$, where $\vartheta(g,p)$ refers to computing the Lov\'asz theta  {function of} a vertex-weighted graph,  {here} $\NOG(H)$ with vertices weighted by probabilities. 
We now invoke the result of Prop.~\ref{prop:sameorthograph} from Appendix \ref{sec:proofs}, namely that all composition methods between the common product and the disjunctive product yield the same non-orthogonality graph, namely the complement to the faithful orthogonality graph introduced in Sec.~\ref{sec:introduceLO}.
Given that those composition rules share the same (non-)orthogonality graph, it follows that all such composition methods allow precisely the same set of $\cQ_1$ models. 
\end{sloppypar}

Consequently, to positively illustrate nontrivial product-choice dependence for $\cQ_1$ models we focus our efforts on finding a difference between the $\cQ_1$ models allowed by the minimal FR-product $\cQ_1(\cB_{3,2,2}^{\min})$ and those allowed by the common FR-product $\cQ_1(\cB_{3,2,2}^{\comm})$.

\subsection{Inequivalence of $\cQ_1$ Models from Minimal to Common FR-Products}\label{sec:varianceofQ1}

A novel result is that $\cQ_1(\mathcal{B}_{n,m,d}^{\min})$ allows strictly more $\cQ_1$ models than the other composition protocols. We demonstrate the observational inequivalence of $\cQ_1(\mathcal{B}_{3,2,2}^{\min})$ and $\cQ_1(\cB_{3,2,2}^{\comm})$ by maximizing various linear functions of conditional probabilities over both {sets}, with the proof hinging on finding functions where the maximum over $\cQ_1(\mathcal{B}_{3,2,2}^{\min})$ is strictly larger than the maximum of $\cQ_1(\mathcal{B}_{3,2,2}^{\comm})$. We limit our attention to linear (as opposed to polynomial) functions, because linear functions are especially amenable to optimization via semidefinite programming (SDP). 

Semidefinite programming powers all the results presented below, because we leveraged the characterization of $\cQ_1$ models in terms of a semidefinite program given in Ref.~\cite[Def.~6.1.2]{AFLS} for all our computational maximization tasks. Note that the SDP characterization is equivalent {to} Def.~\ref{q1mdef} here, as shown in Prop.~6.3.1 of Ref.~\cite{AFLS}, and also equivalent to the definition of $\cQ_1$ models in terms of the Lov\'asz theta number of weighted graphs mentioned in Sec. \ref{sec:equivalenceabovecommon}, as shown in Prop.~6.3.2 of Ref.~\cite{AFLS}.

As a first illustration, we contrast the maximization of the Guess Your-Neighbour's-Input linear function\footnote{Maximizing a linear function is equivalent to {determining} the maximum violation of a linear inequality.} \cite{GYNI} by $\cQ_1$ models as admitted by the $\cB_{3,2,2}^{\min}$ hypergraph versus as admitted by the $\cB_{3,2,2}^{\comm}$ hypergraph. 
While the supremum of this linear function is unity in $\cQ_1(\cB_{3,2,2}^{\comm})$, we find that in $\cQ_1(\mathcal{B}_{3,2,2}^{\min})$ {it} can be maximized further, up to $1.15636$.

The Guess Your-Neighbour's-Input inequality is not the only inequality one can use to show that $\cQ_1(\mathcal{B}_{3,2,2}^{\min})\neq\cQ_1(\cB_{3,2,2}^{\comm})$. {We probe} the entire set of 46 tripartite Bell inequalities representing every symmetry class of facets defining the local polytope in the $(3,2,2)$ Bell scenario\footnote{The local polytope, i.e. the set of classical probabilistic models, may be defined regardless of the choice of composition rule, per Eq.~\eqref{eq:classicalequivalence}.} introduced in Ref. \cite{SLIWA}. These 46 inequalities have a rich history as benchmarks for exploring sets of probabilistic models \cite{SLIWA,JV,PBS}. We find that while most of the 46 inequalities yield the same maximum over both $\cQ_1(\cB_{3,2,2}^{\min})$ and $\cQ_1(\cB_{3,2,2}^{\comm})$, several do not. {Table \ref{ta:sliwa} presents those inequalities which witness $\cQ_1(\cB_{3,2,2}^{\min})\neq\cQ_1(\cB_{3,2,2}^{\comm})$. }

\begin{table}
\centering
\begin{tabular}{c|c|c|c|c|c}
\# ineq & $\cC$ & $\cQ$ & $\cQ_1(^{\comm}\!\otimes)$ & $\cQ_1(^{\min}\!\otimes)$ & $\cG$ \\ 
\hline
 10 & 4 & 4 & {4} & {4.62546} & $\frac{20}{3} =  6.6666...$ \\
 23 & 4& 
 4.68466 & 4.77536 & 5.13836 & 8\\
 25 & 5 & 6.82426 & 6.82426 & 6.83125 & $\frac{31}{3} =  10.3333...$\\
 31 & 6 & 7.80427 & 7.80427 & 7.89009 & 12\\
 32 & 6 & 8.15156 & 8.15156 & 8.15231 & 12\\
 35 & 6 & 7.85528 & 7.85528 & 7.94176 & 12\\
 41 & 7 & 10.36775 & 10.37347 & 10.41218 & 15 \\
 46 & 10 & 12.98520 & 12.98520 & 13.25619 & $\frac{62}{3} =  20.6666...$\\
\hline
\end{tabular}
\caption{Bounds taken from the 46 facets of the $(3,2,2)$ Local Polytope \cite{SLIWA} including maximization over $\cQ_1(\mathcal{B}_{3,2,2}^{\min})$ and $\cQ_1(\cB_{3,2,2}^{\comm})$. In general, we always have that ${\cC \leq \cQ \leq \cQ_1(^{\comm}\!\otimes) \leq \cQ_1(^{\min}\!\otimes) \leq \cG}$. We find that inequalities $\#$ 10, 23, 25, 31, 32, 35, 41, and 46 are those useful for showing that $\cQ_1(^{\min}\!\otimes)$ strictly contains $\cQ_1(^{\comm}\!\otimes)$. As noted in Ref. \cite{JV}, only inequalities $\#23$ and $\#41$ show a gap between the quantum set and  $\cQ_1(^{\comm}\!\otimes)$. For all the remaining inequalities, the bounds over $\cQ$, $\cQ_1(^{\min}\!\otimes)$, and $\cQ_1(^{\comm}\!\otimes)$ coincide. Inequality $\#10$ is just a rescaled version of the Guess Your-Neighbour's-Input inequality \cite{GYNI}.}
\label{ta:sliwa}
\end{table}

Recall that the set of $\cQ_1$ addmited in composition scenarios resulting from the common FR-product are identically the "Almost-Quantum-Correlations" \cite{AQ}, as discussed at length in Sec. 6.4 of Ref. {AFLS}. Accordingly, the bounds appearing in Table \ref{ta:sliwa} may all be found in Ref.~\cite{JV} (which studied tripartite Almost Quantum Correlations) excepting  {those} pertaining to $\cQ_1(\mathcal{B}_{3,2,2}^{\min})$. The classical and GPT bounds for {$\cC$} and $\cG$ respectively are easily computed analytically, using linear programming. The quantum bounds are calculated by optimizing over projective measurements on qubits\footnote{The Bell operator can be encoded as a measurement operator, such that the quantum maximum is given by the largest-possible eigenvalue. Each party's pair of measurement bases can be distinguished by a single real parameter, i.e. the Bloch-sphere angle between the two measurement choices. The computational task is therefore eigenvalue maximization of a matrix with three free variables, and is easily tractable numerically to arbitrary precision. See Ref. \cite{WolfeNewQBs} for further details.} without loss of generality \cite{masanes}. 

This case study of $\cB_{3,2,2}^{\mathrm{prod}}$ shows that sometimes $\cQ_1({H}) \supsetneq \cQ_1(\overbar{H})$, because $\cQ_1(\mathcal{B}_{3,2,2}^{\min})$ was found to be strictly larger than $\cQ_1(\overbar{\cB_{3,2,2}})$, and $\overbar{\cB_{3,2,2}}$ is -- by definition -- the completion of $\mathcal{B}_{3,2,2}^{\min}$. {We intuit} that this is the case in general, i.e. we speculate the following:
\begin{samepage}\begin{conj}
$\cQ_1(H_1)\stackrel{\text{}}{\neq} \cQ_1(H_2)$ whenever $\OG(H_1)\neq \OG(H_2)$, even if $\overbar{H_1}=\overbar{H_2}$. In other words, two contextuality scenarios should admit different $\cQ_1$ models if the two scenario do not have the same associated orthogonality graph, even if the two scenarios are equivalent under completion.
\end{conj}\end{samepage}

Further evidence for this conjecture is presented in in Appendix~\ref{ap:q1comp}, where we contrast a contextuality scenario consisting of $8$ hyperedges with its completion, which has $12$ hyperedges, such that the two hypergraphs have different orthogonality graphs. We show that for the Bell inequality \eqref{eq:ineqAQ}, correlations in $\cQ_1(\overbar{H})$ achieve a maximum value of $1.02325$, whereas those in $\cQ_1(H)$ reach $1.15324$. 

\section{$\cQ_1$ models and Local Orthogonality}\label{se:diss}

In the previous section we have noted that although different composition methods are observationally equivalent as far as classical, quantum, and general probabilistic models are concerned, nevertheless $\cQ_1$ probabilistic models are \emph{not} observationally invariant between some of these compositions. We now discuss how this dependence of $\cQ_1$ models on the choice composition rules is related to Local Orthogonality. 

Per.~Ref.~\cite[Prop. C.2.5]{AFLS}, whenever a composite scenario $H'_{1 \ldots n}$ has a hyperedge set subsuming the common product, i.e. whenever $E(H'_{1 \ldots n}) \supseteq E(H^{\comm}_{1 \ldots n})$, then the composition is guaranteed to capture Local Orthogonality, using the terminology introduced in Sec.~\ref{sec:introduceLO}.

On the other hand, the composite scenarios that arise from the minimal FR-product generally do \emph{not} capture LO\footnote{
The minimal FR-product is not the only composition rule that fails to capture LO. One can think of intermediate composite scenarios between  ${}^{\min}\!\otimes$ and ${}^{\comm}\!\otimes$ as follows: each `product' defines its hyperedges as correlated measurements between the parties where some of them can act in parallel. For instance, in  ${}^{\min}\!\otimes$ the first $n-1$ act in parallel, whereas in ${}^{\comm}\!\otimes$ no two parties act in parallel. Composition rules wherein `some parties act in parallel' define intermediate scenarios which --- like ${}^{\min}\!\otimes$ --- fail to capture LO. The reason for the failure follows the same argument as in ${}^{\min}\!\otimes$: LO includes orthogonal relations such as ${v\underaccent{\mathrm{LO}}{\bot}w}$ where $w$ differs vastly from $v$ in that \emph{only one} party has performed the same local measurement (getting a different outcome), while all other parties perform different measurements in  $w$ relative to $v$. Composition rules where at least two parties act in parallel in every would not include both $v$ and $w$ in any single hyperedge.}. 
To see this,  {let us} briefly review the definition of minimal product pursuant to Def.~\ref{def:minprod} in Appendix~\ref{ap:multiprod}. A hyperedge in ${}^{\min}\!\otimes_k H_k$ may be thought of as a protocol where all but one of the parties (say, the $k^{th}$) choose a measurement $e_i \in E(H_i)$, and the remaining party  {chooses their} measurement $e_k \in E(H_k)$ as a function of the outcome of the other parties.

Consequently, the set of events common to any hyperedge of ${}^{\min}\!\otimes_k H_k$ cannot differ in their multipartite measurement choices by more than variation of a single party's measurement choice. 
On the other hand, a pair of events with only a \emph{single} measurement choice in common can be locally orthogonal. Hence, the minimal product cannot impose all the LO {orthogonality} constraints for scenarios composed by more than two parties. For example, the minimal product $\mathcal{B}_{3,2,2}^{\min}$ cannot capture the local orthogonality  {relation} {${\biggl(\!\begin{smallmatrix}0|0\\0|0\\0|0\end{smallmatrix}\!\biggr)\underaccent{\mathrm{LO}}{\bot}\biggl(\!\begin{smallmatrix}1|0\\0|1\\0|1\end{smallmatrix}\!\biggr)}$}. 

The failure of ${}^{\min}\!\otimes$ to capture LO provides further insight on why $\cQ_1$ models vary relative to the choice of composition rule. This can be seen from the alternative characterization of $\cQ_1$ purely in terms of the non-orthogonality graph of $H$, discussed in the previous section. 

\clearpage
\subsection{Towards a Sensible Composition Rule}\label{se:mfp}

The work of Ref. \cite{AFLS} focused on those composition protocols wherein the resulting composite scenario allows only for probabilistic models that respect the No Signalling principle. Here we argue that, in addition, one should ask the composition rule to capture {the orthogonality relations defined by} Local Orthogonality. 

First, {notice} that the local orthogonality of a pair of events in a composite scenario is an \emph{objective {operational} fact}, having nothing to do with the nature of the composition rule employed. Rather, local orthogonality is a natural way of saying that two events represent contradictory counterfactual possibilities simply by virtue of any counterfactual conflict, regardless of where the conflict is localized. For example, surely the two events {$\biggl(\!\begin{smallmatrix}0|0\\0|0\\0|0\end{smallmatrix}\!\biggr)$ and $\biggl(\!\begin{smallmatrix}1|0\\0|1\\0|1\end{smallmatrix}\!\biggr)$} represent contradictory counterfactual possibilities, because in the first event Alice obtains outcome $0$ for measurement choice $0$, but in the second event Alice obtains the (contradictory) outcome $1$ for the same measurement choice $0$. 

Since orthogonality of events is encoded in the language of composite-scenario hyperedges, it is only natural to ask that sensible composition rules ensure that events which are objectively contradictory counterfactual possibilities (by virtue of Local Orthogonality) are recognized as appropriately orthogonal in the composite scenario. As we have seen, however, the minimal FR-product fails to capture LO.

We argue that an even stronger condition should be met, namely that events should be orthogonal in the composite scenario if and \emph{only} if the events are also locally orthogonal, i.e. the protocol ought to faithfully capture LO. Prop.~\ref{prop:sameorthograph} in Appendix~\ref{sec:proofs} informs us that $^{\comm}\!\otimes$, $^{\max}\!\otimes$, and $^{\disj}\!\otimes$ meet this standard. Indeed, those three composition rules are the \emph{only} candidates the authors are aware of which (1) are associative, (2) capture No Signalling, and (3) faithfully capture Local Orthogonality.

Per Section~\ref{sec:equivalenceabovecommon}, it is clear that the set of $\cQ_1$ models is invariant relative to all composition rules which faithfully capture LO. 

Which multipartite FR-product rules among those which faithfully capture LO) emerges as the most ideal? Well, if one desires that the composite scenario should be self-complete, then this suggests working with the disjunctive protocol. If, however, minimality by hyperedge count is prioritized over near-self-completeness, then one would prefer the common FR-product. We are not aware, {however,} of any compelling reason to prioritize one {choice} over the other.

\section*{Conclusions}

In this work we explore different ways to compose multipartite contextuality scenarios, and the sets of correlations allowed by the different composition rules. Classical, quantum and general probabilistic models are invariant under the choice of composition rule  {within this family}, but we have seen that this invariance does not extend to all other sets of models, as illustrated by $\cQ_1$ models behaving {differently} under $^{\min}\!\otimes$ than under $^{\comm}\!\otimes$. 

The dependence of $\cQ_1$ models on composition rule disappears, however, when we restrict our attention to the subset of composition rules which capture Local Orthogonality. This not only highlights the importance of LO, but arguably suggests a more correct way to define the composition of multipartite contextuality scenarios. 

\section*{Acknowledgements}
We thank Tobias Fritz for fruitful discussions and comments. We also thank an anonymous referee for their valuable comments, from which we substantially improved this manuscript. This research was supported by Perimeter Institute for Theoretical Physics. Research at Perimeter Institute is supported by the Government of Canada through the Department of Innovation, Science and Economic Development Canada and by the Province of Ontario through the Ministry of Research, Innovation and Science.

\appendix\clearpage

\section{Probabilistic Models}\label{ap:models}

Several sets of probabilistic models have {caught} considerable attention \cite{AFLS}, including the set of quantum models and the set of classical models. Each set is defined by a rule for assigning probabilities based on the nature of the underlying supposed physical system. We do not aim to provide a full review in these notes. Instead, we present compact definitions of the two important sets not defined in the main text.

\begin{defn}\label{classical} \textbf{Classical models} \cite[4.1.1]{AFLS}\\
Let $H$ be a contextuality scenario. An assignment of probabilities $p: V(H)\to [0,1]$ is a \emph{classical model} if and only if it can be written as 
\begin{equation}\label{eq:clas}
p(v) = \sum_\lambda q_\lambda p_\lambda(v),
\end{equation}
where the weights $q_\lambda$ and deterministic models $p_\lambda \in \cG(H)$ satisfy 
\begin{equation}
\sum_\lambda q_\lambda = 1 \qquad\text{and}\qquad p_\lambda(v)=\{0,1\} \quad \forall \, v,\lambda.
\end{equation}
The set of all these models is denoted $\cC(H)$.
\end{defn}

\begin{defn}\label{qmdef} \textbf{Quantum models} \cite[5.1.1]{AFLS} \\
Let $H$ be a contextuality scenario. An assignment of probabilities $p: V(H)\to [0,1]$ is a quantum model
if and only if there exists a Hilbert space $\cH$ upon which live some positive-semidefinite quantum state $\rho$ and positive-semidefinite projection operators $\hat{P}_v$ associated to every $v\in V$ such that
\begin{equation}
\label{qmeas}
1=\mathrm{tr}\left( \rho \right),\;\;\;\;p(v) = \mathrm{tr}\left( \rho \hat{P}_v \right) \;\forall v\in V(H),\;\;\textrm{and}\;\;\;\;\sum_{v\in e} \hat{P}_v = \mathbbm{1}_{\mathcal{H}} \;\forall e\in E(H).
\end{equation}
The set of all quantum models is the \emph{quantum set}, denoted $\mathcal{Q}(H)$.
\end{defn}

In the case of Bell scenarios, this definition accords with the usual one of quantum correlations \cite{AFLS}, meaning that each global measurement represented by the projectors $\{\hat{P}_v\}_{v \in e}$ can consistently be expressed as a product of local projectors, one for each party, such that the projectors for different parties commute and are properly normalised. For instance, in a bipartite Bell scenario $\hat{P}_{ab | xy} = \hat{P}_{a|x} \hat{P}_{b|y}$, where $[\hat{P}_{a|x}, \hat{P}_{b|y}]=0$ for all $a,b,x,y$ and $\sum_{a} \hat{P}_{a|x} = \mathbbm{1}_{\mathcal{H}}$ (similarly $\sum_{b} \hat{P}_{b|y} = \mathbbm{1}_{\mathcal{H}}$).

\section{The Foulis-Randall Product of 2 Contextuality Scenarios}\label{ap:prod}

Here we review the binary composition rule suggested by  Foulis and Randall \cite{FR}.

\begin{samepage}
\begin{defn}\cite{FR} \label{FR-prod}
The \emph{Foulis-Randall product} (\emph{FR-product})  {scenario $H_A{^{\FR}\otimes} H_B$ is that} with vertex set ${V(H_A\otimes H_B) = V(H_A)\times V(H_B)}$ and hyperedge set \linebreak ${E(H_A\otimes H_B) = E_{A\rightarrow B} \cup E_{A\leftarrow B}}$ where
\begin{align}
 E_{A\rightarrow B} & \defin \left\{\: \smashoperator{\bigcup_{a\in e_A}}{}\: \{a\} \times f(a) \::\: e_A\in E(H_A),\: f:e_A\to E(H_B)\right\} , \label{eq:FR1}\\[4pt]
 E_{A\leftarrow B} & \defin \left\{\: \smashoperator{\bigcup_{b\in e_B}}{}\: f(b) \times \{b\} \::\: e_B\in E(H_B),\: f:e_B\to E(H_A)\right\} .\label{eq:FR2}
\end{align}
\end{defn}\end{samepage}

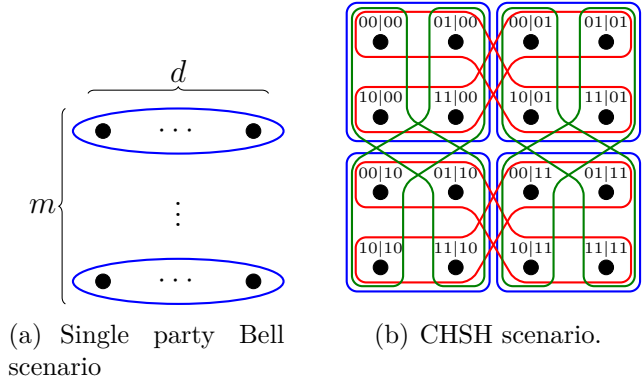
\begin{figure}[tb]
\definecolor{darkgreen}{rgb}{0,.5,0}
\begin{center}
\subfigure[Single party Bell scenario]{
\label{singleparty}
\begin{tikzpicture}
\node[draw,shape=circle,fill,scale=.5] at (0,0) {} ;
\node[draw,shape=circle,fill,scale=.5] at (2,0) {} ;
\node[draw,shape=circle,fill,scale=.5] at (0,2) {} ;
\node[draw,shape=circle,fill,scale=.5] at (2,2) {} ;
\node at (1,0) {$\cdots$} ;
\node at (1,2) {$\cdots$} ;
\draw[thick,blue] (1,0) ellipse (1.4cm and .3cm) ;
\draw[thick,blue] (1,2) ellipse (1.4cm and .3cm) ;
\draw[decoration={brace},decorate] (-.2,2.5) -- (2.2,2.5) ;
\node at (1,2.8) {$d$} ;
\draw[decoration={brace},decorate] (-.5,-.3) -- (-.5,2.3) ;
\node at (-.8,1) {$m$} ;
\node at (1,1) {$\vdots$} ;
\end{tikzpicture}}
\subfigure[CHSH scenario.]{
\label{CHSHc}
\begin{tikzpicture}
\clip (0,.5) rectangle (5,5);
\foreach \x in {0,1} \foreach \y in {0,1} \foreach \a in {0,1} \foreach \b in {0,1}
{
	\node[draw,shape=circle,fill,scale=.5] (e) at (\b+2*\y+1,4-\a-2*\x) {} ;
	\node[above=0pt] at (e) {\tiny{$\a\b|\x\y$}} ;
}
\foreach \x in {0,2} \foreach \y in {0,2} \draw[rounded corners,thick,blue] (\x+0.55,\y+0.68) rectangle (\x+2.45,\y+2.52) ;
\foreach \y in {0,2} \draw[thick,red,rounded corners] (2.8,\y+0.8) -- (4.33,\y+0.8) -- (4.33,\y+1.4) -- (2.8,\y+1.4) -- (2.25,\y+2.4) -- (0.67,\y+2.4) -- (0.67,\y+1.8) -- (2.2,\y+1.8) -- cycle ;
\foreach \y in {0,2} \draw[thick,red,rounded corners] (2.8,\y+1.8) -- (4.33,\y+1.8) -- (4.33,\y+2.4) -- (2.75,\y+2.4) -- (2.2,\y+1.4) -- (0.67,\y+1.4) -- (0.67,\y+0.8) -- (2.2,\y+0.8) -- cycle ;
\foreach \x in {0,2} \draw[thick,darkgreen,rounded corners] (\x+0.62,2.4) -- (\x+0.62,0.75) -- (\x+1.3,0.75) -- (\x+1.3,2.1) -- (\x+2.38,2.8) -- (\x+2.38,4.45) -- (\x+1.65,4.45) -- (\x+1.65,3) -- cycle ;
\foreach \x in {0,2} \draw[thick,darkgreen,rounded corners] (5-\x-0.62,2.4) -- (5-\x-0.62,0.75) -- (5-\x-1.3,0.75) -- (5-\x-1.3,2.1) -- (5-\x-2.38,2.8) -- (5-\x-2.38,4.45) -- (5-\x-1.65,4.45) -- (5-\x-1.65,3) -- cycle ;
\end{tikzpicture}}
\end{center}
\caption{(a) The contextuality scenario $B_{1,m,d}$, a `Bell scenario' with only one party \cite[Fig.~5]{AFLS}. (b) Foulis-Randall product: the CHSH scenario $B_{2,2,2}=B_{1,2,2}{^{\FR}\otimes} B_{1,2,2}$ \cite[Fig.~7]{AFLS}.}
\end{figure}

As an example, consider the situation where we have two parties, Alice and Bob, each  {performing} space-like separated local actions on their share of a system. Locally, each party is characterised by the contextuality scenario of Fig.~\ref{singleparty}, i.e. a hypergraph consisting of $m$ non-intersecting hyperedges (where $m$ is the number of possible measurements  {each party chooses} from) each encompassing $d$ vertices (i.e. the number of outcomes each measurement has). If we denote each such scenario by $H_A$ and $H_B$, the hypergraph to represent the joint situation is $H_{AB} = H_A{^{\FR}\otimes} H_B$ (see Fig.~\ref{CHSHc}). At first one could think such a choice to be arbitrary, but actually it is this choice of product that guarantees that the probabilistic models allowed in the joint scenario $H_{AB}$ satisfy NS; in other words, the set of probabilistic models $\mathcal{G}(H_{AB})$ coincides with the bipartite nonsignalling polytope. 

Indeed, constraint \eqref{eq:FR1} imposes NS from Bob to Alice, and constraint \eqref{eq:FR2} NS from Alice to Bob. For a full explanation of why this is so, see Prop.~3.1.4 of \cite{AFLS}. Here, we will discuss the idea with an example. Consider the CHSH scenario of Fig.~\ref{CHSHc}. 
There, both $\Bigl\lbrace \bigl(\!\begin{smallmatrix}0|0\\0|0\end{smallmatrix}\!\bigr), \bigl(\!\begin{smallmatrix}0|0\\1|0\end{smallmatrix}\!\bigr), \bigl(\!\begin{smallmatrix}1|0\\0|0\end{smallmatrix}\!\bigr), \bigl(\!\begin{smallmatrix}1|0\\1|0\end{smallmatrix}\!\bigr)\Bigr\rbrace$ and $\Bigl\lbrace \bigl(\!\begin{smallmatrix}0|0\\0|0\end{smallmatrix}\!\bigr), \bigl(\!\begin{smallmatrix}0|0\\1|0\end{smallmatrix}\!\bigr), \bigl(\!\begin{smallmatrix}1|0\\0|1\end{smallmatrix}\!\bigr), \bigl(\!\begin{smallmatrix}1|0\\1|1\end{smallmatrix}\!\bigr)\Bigr\rbrace$ are hyperedges of the set given by \eqref{eq:FR1}. In the latter, Bob chooses hyperedge $e_B = \{(0|0), (1|0)\}$ when Alice's outcome is $(0|0)$, and hyperedge $e^\prime_B = \{(0|1), (1|1)\}$ when her outcome is $(1|0)$. Now, since probabilistic models are normalized for each hyperedge, it follows that $p\bigl(\!\begin{smallmatrix}1|0\\0|0\end{smallmatrix}\!\bigr)+p\bigl(\!\begin{smallmatrix}1|0\\1|0\end{smallmatrix}\!\bigr)= p\bigl(\!\begin{smallmatrix}1|0\\0|1\end{smallmatrix}\!\bigr)+ \bigl(\!\begin{smallmatrix}1|0\\1|1\end{smallmatrix}\!\bigr)$, 
i.e.~that Alice's marginal distribution $p_A(1|0)$ does not depend on Bob's choice of measurement. 

\section{Composition Protocols for Multipartite Contextuality Scenarios}\label{ap:multiprod}

The FR-product, however, is not associative, and hence cannot be used to tackle the problem of multipartite composition of contextuality scenarios. Different multipartite extensions were explored in \cite{AFLS}, and in what follows we review three {of them:} The \textit{minimal} FR-product, the \textit{common} FR-product, and the \textit{maximal} FR-product. We also introduce a new composition protocol {that} we call the \textit{disjunctive} FR-product.

\begin{samepage}\begin{defn} \textbf{Minimal FR-product} \cite[3.3.2]{AFLS} \label{def:minprod} \\
The \emph{minimal Foulis-Randall product} $^{\min}\!\otimes_{i=1}^n H_i$ has vertices
\[
V\left(^{\min}\!\otimes_{i=1}^n H_i\right) \defin \prod_i V(H_i) = V(H_1)\times\ldots\times V(H_n),
\]
and $\bigcup_k E_k$ as its set of edges, where the elements of $E_k$ indexed by party $k=1,\ldots,n$ are of the form
\begin{equation}
\label{nFRedge}
\bigcup_{\vec{v}'}\: \{\vec{v}'\} \times f(\vec{v}'),
\end{equation}
where $\vec{v}'$ ranges over $\prod_{i\neq k} e_i$, $e_i\in E(H_i)$ is a choice of edge for every party $i\neq k$, and $\vec{v}\mapsto f(\vec{v})$ is a function which assigns to every joint outcome $\vec{v}=(v_1,\ldots,\bcancel{v_k},\ldots,v_n)$ of all parties except $k$ an edge $f(\vec{v})\in E(H_k)$.
\end{defn}\end{samepage}
 {One such measurement in $E_k$} can be interpreted as follows: each party $i\neq k$ starts by conducting their measurement $e_i$. These parties then announce their joint outcome $\vec{v}'$ to the remaining party $k$, who conducts a measurement $f(\vec{v}')$ chosen as a function of the previous joint outcome. 
These hyperedges, hence, impose No Signalling constraints from any single party to the rest. 

The minimal product may also be described in an intuitive way, in terms of simple scenarios that we define in the following. First, given a set of contextuality scenarios $\{ H_1, \ldots, H_l \}$ one can define their simplest composite scenario $H^{\textrm{sim}}$ as that with $V(H^{\textrm{sim}}) = \prod_i V(H_i)$ and $E(H^{\textrm{sim}}) = \prod_i E(H_i)$, i.e. both the vertex set and the edge set of $H^{\textrm{sim}}$ are the tensor product of the individual vertex and edge sets respectively. This simple composition rule, $^{\textrm{sim}}\otimes$, does not capture No Signalling, and is therefore of limited utility. However, the hyperedges in the minimal product scenario may be understood as elements of set $E_{A\rightarrow B}$ constructed in the definition of the bipartite FR-product, where now $B$ consists of one scenario $H_k$, and $A$ denotes the joint simple scenario $^{\textrm{sim}}\otimes_{\neg k}$ composed by the remaining $n\!-\!1$ scenarios $\{H_1,\ldots, H_n \}\setminus H_k$. The collection of hyperedges that arise from this construction $E_{A\rightarrow B}$, for a particular $k$, {defines} the set $E({^{\textrm{sim}}\otimes_{\neg k} \rightarrow k})$. Hence, 
\begin{align*}
E(^{\min}\!\otimes_{i=1}^n H_i) = \cup_k E({^{\textrm{sim}}\otimes_{\neg k} \rightarrow k})\,.
\end{align*}

\begin{defn} \textbf{Common FR-product} \cite{AFLS} \\
The \emph{common Foulis-Randall product} $^{\comm}\!\otimes_{i=1}^n H_i$ has vertices
\[
V\left(^{\comm}\!\otimes_{i=1}^n H_i\right) \defin \prod_i V(H_i) = V(H_1)\times\ldots\times V(H_n),
\]
and its edge set is the union of the edge sets from all the possible iterated binary Foulis-Randall products among  {$\{H_i\}$.}
\end{defn}
The hyperedges in the common product have an operational interpretation in terms of \textit{correlated measurements}. A correlated measurement is defined by a fixed order of the parties $i_1, \ldots, i_n$, and a function $f_j$ for each party $j$, such that the $k$-th party in the protocol chooses a measurement $e_{i_k} \in E(H_{i_k})$ as a function of the outcomes of the previous parties in the protocol  {(i.e.~$e_{i_k} = f_{i_k}(v_{i_1}, \ldots, v_{i_{k-1}})$)}, obtains an outcome $v_{i_k} \in e_{i_k}$, and communicates it to the remaining $n-k$ parties $i_{k+1}, \ldots i_n$. 

To illustrate the difference between the minimal and the common product scenarios, consider the tripartite Bell scenarios $\mathcal{B}_{3,2,2}^{\min} = {}^{\min}\!\otimes  H_i$ and $\mathcal{B}_{3,2,2}^{\comm} = {}^{\comm}\!\otimes  H_i$, with $H_i = \mathcal{B}_{1,2,2}$  {and $i \in \{1,2,3\}$.} A hyperedge that appears in $\mathcal{B}_{3,2,2}^{\comm}$ is given by:
\begin{align*}
e = \left\lbrace
\biggl(\!\begin{smallmatrix}0|0\\0|0\\0|0\end{smallmatrix}\!\biggr),
\biggl(\!\begin{smallmatrix}0|0\\0|0\\1|0\end{smallmatrix}\!\biggr),
\biggl(\!\begin{smallmatrix}0|0\\1|0\\0|0\end{smallmatrix}\!\biggr),
\biggl(\!\begin{smallmatrix}0|0\\1|0\\1|0\end{smallmatrix}\!\biggr),
\biggl(\!\begin{smallmatrix}1|0\\0|1\\0|0\end{smallmatrix}\!\biggr),
\biggl(\!\begin{smallmatrix}1|0\\0|1\\1|0\end{smallmatrix}\!\biggr),
\biggl(\!\begin{smallmatrix}1|0\\1|1\\0|1\end{smallmatrix}\!\biggr),
\biggl(\!\begin{smallmatrix}1|0\\1|1\\1|1\end{smallmatrix}\!\biggr)\right\rbrace\,.
\end{align*}
There, Alice chooses the measurement $\{(0|0), (1|0)\}$. The next party in the protocol is Bob: if Alice obtains outcome $(0|0)$ he performs measurement $\{(0|0), (1|0)\}$, and if she obtains outcome $(1|0)$ he performs measurement $\{(0|1), (1|1)\}$. Finally, Charlie chooses and performs a measurement: he chooses measurement $\{(0|0), (1|0)\}$ unless both Alice's outcome is $(1|0)$ and Bob's is $(1|1)$,  {in which case he chooses $\{(0|1), (1|1)\}$. }

Notice that the hyperedge $e$ cannot belong to $E(\mathcal{B}_{3,2,2}^{\min})$. This is because more than one party's measurement choice  {vary} among the events in $e$. 

\bigskip
Finally, before presenting the maximal product, we need to introduce the notion of a measurement protocol \cite{AFLS, LO2}:

\begin{defn} \textbf{Measurement Protocol} \cite[3.3.4]{AFLS} \\
Let $S$ be a set of parties. A \emph{measurement protocol} $\mathcal{P}$ for $S$ consists of the following data:
\begin{enumerate}
\item if $S=\emptyset$, the unique protocol is $\mathcal{P}=\emptyset$,
\item\label{measprotb} otherwise, the protocol is a triple $\mathcal{P}=(k,e,f)$, where $k\in S$ is a party, $e\in E(H_k)$ is an edge, and $f$ is a function assigning to each vertex $v\in e$ a measurement protocol $f(v)$ on $S\setminus\{k\}$.
\end{enumerate}
\end{defn}
An outcome of a measurement protocol $\mathcal{P}$ for all parties $S$ has exactly one component in each $V(H_i)$ for each $i\in S$, so that it can be regarded as an element of $\prod_{i\in S} V(H_i)$.

Now we can define the maximal FR-product:
\begin{samepage}\begin{defn} \textbf{Maximal FR-product} \cite[3.3.6]{AFLS} \\
The \emph{maximal Foulis-Randall product} $^{\max}\!\otimes_{i=1}^n H_i$ has vertices $
V\left(^{\max}\!\otimes_{i=1}^n H_i\right) \defin \prod_i V(H_i)$ and set of edges $\bigcup_{\mathcal{P}} O(\mathcal{P})$ where $\mathcal{P}$ is a measurement protocol for $\{1, \ldots, n\}$,  {and $O(\mathcal{P})$ denotes the outcome set of $\mathcal{P}$.}
\end{defn}\end{samepage}
Operationally, each hyperedge in $^{\max}\!\otimes$ has the form of a correlated measurement where the order of the parties is adapted dynamically throughout the protocol \cite{LO2}. 

To illustrate the difference between the common and the maximal product scenarios, consider the tripartite Bell scenarios $\mathcal{B}_{3,2,2}^{\comm}$ and $\mathcal{B}_{3,2,2}^{\max} = {}^{\max}\!\otimes  H_i$. 
A hyperedge that appears in $\mathcal{B}_{3,2,2}^{\max}$ is given by:
\begin{align*}
e = \left\lbrace
\biggl(\!\begin{smallmatrix}0|0\\0|0\\0|1\end{smallmatrix}\!\biggr),
\biggl(\!\begin{smallmatrix}0|0\\0|0\\1|1\end{smallmatrix}\!\biggr),
\biggl(\!\begin{smallmatrix}0|0\\1|0\\0|0\end{smallmatrix}\!\biggr),
\biggl(\!\begin{smallmatrix}0|0\\1|0\\1|0\end{smallmatrix}\!\biggr),
\biggl(\!\begin{smallmatrix}1|0\\0|1\\0|1\end{smallmatrix}\!\biggr),
\biggl(\!\begin{smallmatrix}1|0\\1|1\\0|1\end{smallmatrix}\!\biggr),
\biggl(\!\begin{smallmatrix}1|0\\0|0\\1|1\end{smallmatrix}\!\biggr),
\biggl(\!\begin{smallmatrix}1|0\\1|0\\1|1\end{smallmatrix}\!\biggr)\right\rbrace\,.
\end{align*}
There, Alice initiates the protocol by choosing the measurement $\{(0|0), (1|0)\}$. 
\begin{compactitem}
\item[-] When her outcome is $(0|0)$, the next party in the protocol is Bob, and he chooses the hyperedge $\{(0|0), (1|0)\}$. The third party in the protocol is then Charlie. When Bob's outcome is $(0|0)$ he chooses the hyperedge $\{(0|1), (1|1)\}$, and otherwise $\{(0|0), (1|0)\}$. 
\item[-] When Alice's outcome is $(1|0)$, the second party in the protocol is Charlie, and he measures $\{(0|1), (1|1)\}$. The third party in the protocol is then Bob. When Charlie's outcome is $(0|1)$ he chooses the hyperedge $\{(0|1), (1|1)\}$, and otherwise $\{(0|0), (1|0)\}$. 
\end{compactitem}
This hyperedge $e$ does not belong to $E(\mathcal{B}_{3,2,2}^{\comm})$, since it cannot arise from a measurement protocol where the order of the parties is fixed.

\bigskip
The disjunctive FR-product is motivated by Local Orthogonality  {relations (see Sec.~\ref{sec:introduceLO}),} and is meant to be a generalization the disjunctive graphical product introduced in Sec.~\ref{sec:introduceLO} to a product rule for hypegraphs. The disjunctive hypergraph FR-product is constructed as follows: First, construct the faithful orthgonality graph using the standard disjunctive graphical product per Eq.~\eqref{eq:minLOgraph}, and identify all cliques of mutually-adjacent vertices in the faithful orthogonality graph. Each such clique defines a candidate hyperedge $e$. Next, we filter this candidate pool by accepting a candidate hyperedge into the composite scenario if and only if that hyperedge is a linear combination of the normalization hyperedges and the No Signalling equalities, as defined by the normalization and No Signalling matrices introduced in Section \ref{sec:compositionprereq} of the main text.
\begin{samepage}\begin{defn} \textbf{Disjunctive FR-product}\label{def:disj} \\
 {The \emph{Disjunctive FR-product} $^{\disj}\!\otimes_{i=1}^n H_i$ has vertices $
V\left(^{\max}\!\otimes_{i=1}^n H_i\right) \defin \prod_i V(H_i)$ and set of edges}:
\hypersetup{linkcolor=black}
\begin{align*}
&E\left(^{\disj}\!\otimes_{i=1}^n H_i\right) = \Big\{\quad e \quad : \quad\text{such that both}\\
\hspace{-3.5pt}\text{\ref{def:disj}a.}\quad& e\in\operatorname{Cliques}\left(\operatorname{FaithfulOrthoGraph}(H_1,...H_n)\right)  \\
\hspace{-3.5pt}\text{\ref{def:disj}b.}\quad&\operatorname{\mathtt{rank}}\left(M_{\textrm{Normalization}}^{\vec{\bf{1}}} \cup M_{\textrm{NoSignalling}}^{\vec{\bf{0}}}\right)=\operatorname{\mathtt{rank}}\left(e\cup M_{\textrm{Normalization}}^{\vec{\bf{1}}} \cup M_{\textrm{NoSignalling}}^{\vec{\bf{0}}}\right) \quad 
\Big\}\,.
\end{align*}
\end{defn}\end{samepage}

To illustrate the difference between the scenarios induced by the disjunctive FR-product and the maximal FR-product, consider the tripartite Bell scenarios $\mathcal{B}_{3,2,2}^{\max}$ and $\mathcal{B}_{3,2,2}^{\disj} = {}^{\disj}\!\otimes  H_i$. 
A hyperedge that appears in $\mathcal{B}_{3,2,2}^{\disj}$ is given by:
\begin{align*}
e = \left\lbrace
\biggl(\!\begin{smallmatrix}0|0\\0|0\\0|1\end{smallmatrix}\!\biggr),
\biggl(\!\begin{smallmatrix}1|0\\0|0\\0|1\end{smallmatrix}\!\biggr),
\biggl(\!\begin{smallmatrix}0|1\\1|0\\0|0\end{smallmatrix}\!\biggr),
\biggl(\!\begin{smallmatrix}0|1\\1|0\\1|0\end{smallmatrix}\!\biggr),
\biggl(\!\begin{smallmatrix}0|1\\0|0\\1|1\end{smallmatrix}\!\biggr),
\biggl(\!\begin{smallmatrix}1|1\\1|0\\0|1\end{smallmatrix}\!\biggr),
\biggl(\!\begin{smallmatrix}1|1\\0|1\\1|1\end{smallmatrix}\!\biggr),
\biggl(\!\begin{smallmatrix}1|1\\1|1\\1|1\end{smallmatrix}\!\biggr)\right\rbrace\,.
\end{align*}
This hyperedge $e$ does not belong to $E(\mathcal{B}_{3,2,2}^{\max})$, since there is no party that may act as the initial one in the measurement protocol, i.e. with a fixed choice of hyperedge $e_k \in E(H_k)$.
 
\bigskip
\begin{samepage}
\begin{defn} \textbf{Completion FR-product} \\
The \emph{completion Foulis-Randall product} $\overbar{\otimes}_{i=1}^n H_i$ is defined as the completion of any of the other multipartite FR-products, as those composition rules are known to be equivalent under completion (See Prop.~\ref{prop:equivalentundercompletion} in Appendix~\ref{sec:proofs}). Thus $\overbar{\otimes}_{i=1}^n H_i \equiv \overbar{^{\min}\!\otimes_i H_i}=\overbar{^{\comm}\!\otimes_i H_i}=\overbar{^{\max}\!\otimes_i H_i}=\overbar{^{\disj}\!\otimes_i H_i}\,.$
\end{defn}\end{samepage}

\section{Propositions and Proofs}\label{sec:proofs}
\begin{prop}\label{prop:hierarchyofedges} The hyperedges of the the maximal FR-product are a subset of the hyperedges of the disjunctive FR-product, i.e. $E(^{\max}\!\otimes_i H_i)\subseteq E(^{\disj}\!\otimes_i H_i)$. A trivial corollary taking account the other definitions in Appendix~\ref{ap:multiprod} is that $
E(^{\min}\!\otimes_i H_i) \subseteq
E(^{\comm}\!\otimes_i H_i)\subseteq
E(^{\max}\!\otimes_i H_i)\subseteq 
E(^{\disj}\!\otimes_i H_i) \subseteq
E(\overbar{\otimes_i H_i})\,.$
\begin{proof}We claim that every hyperedge in ${}^{\max}\!\otimes_i H_i$ is a hyperedge in ${}^{\disj}\!\otimes_i H_i$. Per Definition \ref{def:disj}a, this is equivalent to claiming that every hyperedge in ${}^{\max}\!\otimes_i H_i$ is a clique of locally orthogonal events. Recall that the hyperedges of ${}^{\max}\!\otimes_i H_i$ correspond to measurement protocols. Any time we compare distinct endpoints in a measurement protocol we can isolate (at least one) branching point in the protocol to explain the final divergence. Each branching point in a protocol is an instance of Local Orthogonality,  {i.e.~a party performing a local measurement $e_j$ and obtaining different outcomes.} Therefore every event in a hyperedge corresponding to a measurement protocol must be locally orthogonal to every other even in the hyperedge.
\end{proof}\medskip
\end{prop}

\begin{prop}\label{prop:equivalentundercompletion} The disjunctive FR-product has the same completion as the other FR-products, i.e. ${\overbar{^{\min}\!\otimes_i H_i}=\overbar{^{\comm}\!\otimes_i H_i}=\overbar{^{\max}\!\otimes_i H_i}=\overbar{^{\disj}\!\otimes_i H_i}\,.}$
\begin{proof}To show this, we simply need to show that the linear span of the disjunctive FR-product's hyperedges is not a higher dimension than the linear span of the minimal FR-product's hyperedges.  {This suffices since }${\overbar{^{\min}\!\otimes_i H_i}=\overbar{^{\comm}\!\otimes_i H_i}=\overbar{^{\max}\!\otimes_i H_i}}$ was already shown in Ref.~\cite[Appendix C]{AFLS}, and we already know that the disjunctive FR-product induces hypegraphs which contain all hyperedges generated by the other products. To prove ${\mathtt{rref}\!\left(E^{\vec{\bf{1}}}({}^{\disj}\!\otimes_i H_i)\right)}\subseteq{\mathtt{rref}\!\left(E^{\vec{\bf{1}}}({}^{\min}\!\otimes_i H_i)\right)}$ we recall that even the minimal FR-product captures No Signalling, again as shown in, and therefore 
${\mathtt{rref}\!\left(M_{\textrm{Normalization}}^{\vec{\bf{1}}} \!\cup\! M_{\textrm{NoSignalling}}^{\vec{\bf{0}}}\right)}\subseteq{\mathtt{rref}\!\left(E^{\vec{\bf{1}}}({}^{\min}\!\otimes_i H_i)\right)}$. On the other hand, by Definition~\ref{def:disj}b, the linear span of the disjunctive product is bounded by that of Normalization and No Signalling, that is, 
${{\mathtt{rref}\!\left(\!E^{\vec{\bf{1}}}({}^{\disj}\!\otimes_i H_i)\right)} \subseteq\linebreak {\mathtt{rref}\!\left(M_{\textrm{Normalization}}^{\vec{\bf{1}}} \!\cup\! M_{\textrm{NoSignalling}}^{\vec{\bf{0}}}\right)}}$.
\end{proof}\end{prop}
A corollary of Prop~\ref{prop:equivalentundercompletion} is that the maximal FR-product is not self-complete even when all the individual hypergraphs \emph{are} self-complete, , i.e. $^{\max}\!\otimes \overbar{H_i}\neq\overbar{^{\max}\!\otimes_i H_i}$. This is because we know of an instance where ${}^{\max}\!\otimes_i H_i\neq {}^{\disj}\!\otimes_i H_i$ while the individual scenarios are complete, coupled with Prop.~\ref{prop:equivalentundercompletion} which informs us that the disjunctive FR-product is intermediate to the maximal FR-product and its completion. The counterexample is the composition of Bell scenarios considered in Section \ref{se:q1multi}. Indeed, $\overline{\mathcal{B}_{1,m,d}}=\mathcal{B}_{1,m,d}$ as the single-partite scenarios have no intersecting hyperedges, but we found that $\mathcal{B}_{3,2,2}^{\disj}$ has 64 more hyperedges than $\mathcal{B}_{3,2,2}^{\max}$. This  {disproves the following conjecture of Ref.~\citep[Conj.~C.2.6]{AFLS}: $\overbar{\otimes}_{i=1}^n H_i \stackrel{\text{Conj.}}{=} {}^{\max}\!\otimes_{i=1}^n \overbar{H_i}$.}

\medskip

\begin{prop}\label{eq:modelsinvariant} 
The sets of classical and quantum models remain invariant under any choice multipartite FR-product defined in Appendix~\ref{ap:multiprod}, i.e. \linebreak $\cC({}^{\min}\!\otimes_i H_i) = \cC({}^{\comm}\!\otimes_i H_i) = \cC({}^{\max}\!\otimes_i H_i) = \cC({}^{\disj}\!\otimes_i H_i) = \cC(\overbar{\otimes_i H_i})$, and also $\cQ({}^{\min}\!\otimes_i H_i) = \cQ({}^{\comm}\!\otimes_i H_i) = \cQ({}^{\max}\!\otimes_i H_i) = \cQ({}^{\disj}\!\otimes_i H_i) = \cQ(\overbar{\otimes_i H_i})$. (This claim also follows for the set of general probabilistic models by definition.)
\begin{proof}
This follows from Ref.~\cite[Prop. C.1.6]{AFLS} where it was proven that both classical and quantum models are invariant under completion. Working backwards, whatever the classical or quantum models are for $\overbar{\otimes}_{i=1}^n H_i$, they must be the same for all the other composition rules, per Prop.~\ref{prop:equivalentundercompletion}.
\end{proof}\medskip\end{prop}

\begin{sloppypar}
\begin{prop}\label{prop:sameorthograph} The common, maximal and disjunctive FR-products all give rise to precisely the same orthogonality conditions; more specifically, those products faithfully capture Local Orthogonality. Formally, ${\operatorname{FaithfulOrthoGraph}(H_1,...,H_n)} ={  
\OG({^{\comm}\!\otimes_i H_i})} = {\OG({^{\max}\!\otimes_i H_i})} = {\OG({^{\disj}\!\otimes_i H_i})\,.}$
\begin{proof}
The claim is partially justified by Prop.~C.2.5 of \cite{AFLS}, which states that \emph{whenever} two vertices are locally orthogonal then their composite extensions are also orthogonal in ${}^{\comm}\!\otimes$. In other words, it is already known that the common product captures Local Orthogonality; Prop~\ref{prop:sameorthograph} is coming to show that it captures LO \emph{faithfully}. Definition~\ref{def:disj}a ensures that the disjunctive FR-product does not impose any orthogonality relations \emph{beyond} LO. Since Prop.~\ref{prop:hierarchyofedges} plainly guarantees that the orthogonality relations implied by ${^{\comm}\!\otimes_i H_i}$ cannot exceed those implied by ${^{\disj}\!\otimes_i H_i}$, the only conclusion is that the common and disjunctive FR-products both faithfully capture Local Orthogonality, justifying the claim.
\end{proof}\medskip\end{prop}\end{sloppypar}

Finally, we update the now-disproven Conj.~C.2.6 of Ref.~\cite{AFLS} with the following replacement conjecture regarding the completion product:
\begin{samepage}\begin{conj}\label{completionconjecture}
The disjunctive FR-product gives rise to a self-complete composite hypergraph whenever the individual hypergraphs being composed are themselves self-complete, i.e. we conjecture that $
\overbar{\otimes}_{i=1}^n \overbar{H_i} = {}^{\disj}\!\otimes_i \overbar{H_i}\,.$\end{conj}\end{samepage}

\noindent\textit{The intuition behind this conjecture (though not a full proof) is as follows. } 
Let $H = {}^{\disj}\!\otimes_i H_i$. Recall that every virtual hyperedge $e_V$ in $H$ cannot increase the rank of $E^{\vec{\bf{1}}}(H)$ when appending $e_V$ as a row of $E(H)$. Hence, $e_V$ can be though of as coming from a linear combination of the rows of $E(H)$, which contains $M_{\textrm{Normalization}}$ and has the same reduced echelon form as $M_{\textrm{Normalization}} \cup M_{\textrm{NoSignalling}}$. Operationally, then, one can think of the rows of $M_{\textrm{NoSignalling}}$  as substitution rules to `generate' new hyperedges from the existing ones in $E(H)$, i.e.~they allow one to substitute old hypernodes for new ones in existing hyperedges, hence generating new ones. 

Consider a generic hyperedge $ e \in E({}^{\disj}\!\otimes_i H_i)$, which by definition consists of a set of mutually LO vertices. Without loss of generality, let the No Signalling substitution rule $ns \in M_{\textrm{NoSignalling}}$ be $\{v,v',v'',...\}\to\{w,w',w'',...\}$ such that ${e_V=e+n\!s = (e\setminus \{v,v',v'',...\})\cup \{w,w',w'',...\}}$. By the definition of a No Signalling equality, the events $\{v,v',v'',...\}$ which appear on one side of the equality must all be identical up the local outcomes of one party, i.e. generically that subset of vertices has the form $v\mathpunct{=}\left(\!\begin{smallmatrix}v_1\vspace{-1ex}\\\vdots\\v_k\vspace{-1ex}\\\vdots\\v_n\end{smallmatrix}\!\right)$, $v'\mathpunct{=}\left(\!\begin{smallmatrix}v_1\vspace{-1ex}\\\vdots\\{v'}_k\vspace{-1ex}\\\vdots\\v_n\end{smallmatrix}\!\right)$, and $v''\mathpunct{=}\left(\!\begin{smallmatrix}v_1\vspace{-1ex}\\\vdots\\{v''}_k\vspace{-1ex}\\\vdots\\v_n\end{smallmatrix}\!\right)$ etc, 
where $\{v_k,{v'}_k,{v''}_k,...\}$ are all members of a local hyperedge in the contextuality scenario of party $k$. Moreover, without loss of generality we may take $\{v_k,{v'}_k,{v''}_k,...\}$ to be identically some hyperedge in $H_k$, noting that possibly $\{v_k,{v'}_k,{v''}_k,...\}$ intersects with $\{w_k,{w'}_k,{w''}_k,...\}$, in which case application of No Signalling to a hyperedge `substitutes' some nodes for themselves.

Every node in $e\setminus \{v,v',v'',...\}$ is locally orthogonal to all the nodes $\{v,v',v'',...\}$, as this is the assumption of our induction. The nature of the local orthogonality is not arbitrary, however: It \emph{cannot} be that the \emph{only} reason why a node in $e\setminus \{v,v',v'',...\}$ is locally orthogonal to a node in $\{v,v',v'',...\}$ is due to differing outcomes in the local hyperedge $\{v_k,{v'}_k,{v''}_k,...\}$, because otherwise that node would be a \emph{member} of $\{v,v',v'',...\}$. Therefore the localization of the orthogonality between the events inside $\{v,v',v'',...\}$ versus the events of $e\setminus \{v,v',v'',...\}$ must be due to parties other than party $k$. Since $\{w,w',w'',...\}$ is identical to $\{v,v',v'',...\}$ aside from the local events drawn from party $k$, it follows that every node in $e\setminus \{v,v',v'',...\}$ is locally orthogonal to all the nodes $\{w,w',w'',...\}$. 

As such, any hyperedge generated by applying No Signalling cannot introduce novel orthogonality beyond local orthogonality. Formally, if $e_V=e+ n\!s$ where $e$ is some hyperedge consisting of $100\%$ mutually LO events and $n\!s \in M_{\textrm{NoSignalling}}$, and moreover $e_V$ is a virtual hyperedge (in the sense that $e_V$ contains only zeroes and ones when expressed in matrix notation), then  $e_V$ also consists of $100\%$ mutually LO events.

One way to extend this small proof into a full proof of the conjecture would be showing that \emph{every} virtual hyperedge of ${}^{\disj}\!\otimes \overbar{H_i}$ belongs to the following (finite) hierarchy: Starting with level $L_0\equiv M_{\textrm{Normalization}}$, construct subsequent sets of virtual hyperedges by defining level $L_{q+1}$ as all virtual hyperedges which can be expressed as $e_V=e+ n\!s$ where $e\in L_{q}$ and, as usual, $n\!s \in M_{\textrm{NoSignalling}}$. Showing that every virtual hyperedge of ${}^{\disj}\!\otimes_i \overbar{H_i}$ belongs to $L_{q\to\inf}$ would hence constitute a proof of Conj.~\ref{completionconjecture}; assessing whether or not all virtual hyperedges fall into this hierarchy is deferred to future work. 

Finally, we emphasize the importance of the qualifier ``whenever the individual hypergraphs being composed are themselves self-complete" in Conj.~\ref{completionconjecture}. Note that all the virtual hyperedges of an \emph{incomplete} $H_k$ would translate into virtual hyperedges of $M_{\textrm{Normalization}}$ without any reference to $M_{\textrm{NoSignalling}}$, and hence we expect virtual hyperedges beyond those in $L_{q\to\inf}$, i.e. outside the hierarchy built up via successive applications of No Signalling. Indeed, if any individual hypergraph $H_k$ is ortho-unstable, then the composition ${}^{\disj}\!\otimes_i H_i$ is also ortho-unstable, and hence certainly \emph{not} self-complete.

\section{Simple Example that $\cQ_1$ Changes Under Completion}\label{ap:q1comp}

The question that we explore in this section is whether there exists a contextuality scenario $H$ such that $\cQ_1(\overbar{H}) \subsetneq \cQ_1(H)$. Indeed, the findings in the previous section show that this is actually the case, by taking $H = \mathcal{B}_{1,2,2} \, ^{\min}\!\otimes \mathcal{B}_{1,2,2} \, ^{\min}\!\otimes \mathcal{B}_{1,2,2}$. 

We therefore shift our focus to try and find a simpler example of the fact. For that, we consider now a CHSH type of scenario, that is two parties, Alice and Bob, each acting locally on a contextuality scenario $H_A = \mathcal{B}_{1,2,2}$ and $H_B = \mathcal{B}_{1,2,2}$. Now define two joint scenarios as follows. 

Let the first join scenario $H_1(V,E)$ have vertex set ${V(H_1)\defin V(H_A) \times V(H_B)}$ and hyperedge set $E(H_1)\defin$
\begin{align*}
\begin{tabular}{l|cccccccccccccccc}
  & 
\(\!\!\bigl(\!\begin{smallmatrix}
 0|0 \\
 0|0 \\
\end{smallmatrix}\!\bigr)\!\!\)
 & 
\(\!\!\bigl(\!\begin{smallmatrix}
 0|0 \\
 1|0 \\
\end{smallmatrix}\!\bigr)\!\!\)
 & 
\(\!\!\bigl(\!\begin{smallmatrix}
 1|0 \\
 0|0 \\
\end{smallmatrix}\!\bigr)\!\!\)
 & 
\(\!\!\bigl(\!\begin{smallmatrix}
 1|0 \\
 1|0 \\
\end{smallmatrix}\!\bigr)\!\!\)
 & 
\(\!\!\bigl(\!\begin{smallmatrix}
 0|0 \\
 0|1 \\
\end{smallmatrix}\!\bigr)\!\!\)
 & 
\(\!\!\bigl(\!\begin{smallmatrix}
 0|0 \\
 1|1 \\
\end{smallmatrix}\!\bigr)\!\!\)
 & 
\(\!\!\bigl(\!\begin{smallmatrix}
 1|0 \\
 0|1 \\
\end{smallmatrix}\!\bigr)\!\!\)
 & 
\(\!\!\bigl(\!\begin{smallmatrix}
 1|0 \\
 1|1 \\
\end{smallmatrix}\!\bigr)\!\!\)
 & 
\(\!\!\bigl(\!\begin{smallmatrix}
 0|1 \\
 0|0 \\
\end{smallmatrix}\!\bigr)\!\!\)
 & 
\(\!\!\bigl(\!\begin{smallmatrix}
 0|1 \\
 1|0 \\
\end{smallmatrix}\!\bigr)\!\!\)
 & 
\(\!\!\bigl(\!\begin{smallmatrix}
 1|1 \\
 0|0 \\
\end{smallmatrix}\!\bigr)\!\!\)
 & 
\(\!\!\bigl(\!\begin{smallmatrix}
 1|1 \\
 1|0 \\
\end{smallmatrix}\!\bigr)\!\!\)
 & 
\(\!\!\bigl(\!\begin{smallmatrix}
 0|1 \\
 0|1 \\
\end{smallmatrix}\!\bigr)\!\!\)
 & 
\(\!\!\bigl(\!\begin{smallmatrix}
 0|1 \\
 1|1 \\
\end{smallmatrix}\!\bigr)\!\!\)
 & 
\(\!\!\bigl(\!\begin{smallmatrix}
 1|1 \\
 0|1 \\
\end{smallmatrix}\!\bigr)\!\!\)
 & 
\(\!\!\bigl(\!\begin{smallmatrix}
 1|1 \\
 1|1 \\
\end{smallmatrix}\!\bigr)\!\!\)
 \\\hline
 \(\!e_1\!\) & 1 & 1 & 1 & 1 & \({\scriptscriptstyle ^0}\) & \({\scriptscriptstyle ^0}\) & \({\scriptscriptstyle ^0}\) & \({\scriptscriptstyle ^0}\) & \({\scriptscriptstyle ^0}\) & \({\scriptscriptstyle ^0}\) & \({\scriptscriptstyle ^0}\) & \({\scriptscriptstyle ^0}\) & \({\scriptscriptstyle ^0}\) & \({\scriptscriptstyle ^0}\) & \({\scriptscriptstyle ^0}\) & \({\scriptscriptstyle ^0}\) \\
 \(\!e_2\!\) & 1 & 1 & \({\scriptscriptstyle ^0}\) & \({\scriptscriptstyle ^0}\) & \({\scriptscriptstyle ^0}\) & \({\scriptscriptstyle ^0}\) & 1 & 1 & \({\scriptscriptstyle ^0}\) & \({\scriptscriptstyle ^0}\) & \({\scriptscriptstyle ^0}\) & \({\scriptscriptstyle ^0}\) & \({\scriptscriptstyle ^0}\) & \({\scriptscriptstyle ^0}\) & \({\scriptscriptstyle ^0}\) & \({\scriptscriptstyle ^0}\) \\
 \(\!e_3\!\) & 1 & \({\scriptscriptstyle ^0}\) & 1 & \({\scriptscriptstyle ^0}\) & \({\scriptscriptstyle ^0}\) & \({\scriptscriptstyle ^0}\) & \({\scriptscriptstyle ^0}\) & \({\scriptscriptstyle ^0}\) & \({\scriptscriptstyle ^0}\) & 1 & \({\scriptscriptstyle ^0}\) & 1 & \({\scriptscriptstyle ^0}\) & \({\scriptscriptstyle ^0}\) & \({\scriptscriptstyle ^0}\) & \({\scriptscriptstyle ^0}\) \\
 \(\!e_4\!\) & \({\scriptscriptstyle ^0}\) & 1 & \({\scriptscriptstyle ^0}\) & 1 & \({\scriptscriptstyle ^0}\) & \({\scriptscriptstyle ^0}\) & \({\scriptscriptstyle ^0}\) & \({\scriptscriptstyle ^0}\) & 1 & \({\scriptscriptstyle ^0}\) & 1 & \({\scriptscriptstyle ^0}\) & \({\scriptscriptstyle ^0}\) & \({\scriptscriptstyle ^0}\) & \({\scriptscriptstyle ^0}\) & \({\scriptscriptstyle ^0}\) \\
 \(\!e_5\!\) & \({\scriptscriptstyle ^0}\) & \({\scriptscriptstyle ^0}\) & 1 & 1 & 1 & 1 & \({\scriptscriptstyle ^0}\) & \({\scriptscriptstyle ^0}\) & \({\scriptscriptstyle ^0}\) & \({\scriptscriptstyle ^0}\) & \({\scriptscriptstyle ^0}\) & \({\scriptscriptstyle ^0}\) & \({\scriptscriptstyle ^0}\) & \({\scriptscriptstyle ^0}\) & \({\scriptscriptstyle ^0}\) & \({\scriptscriptstyle ^0}\) \\
 \(\!e_6\!\) & \({\scriptscriptstyle ^0}\) & \({\scriptscriptstyle ^0}\) & \({\scriptscriptstyle ^0}\) & \({\scriptscriptstyle ^0}\) & 1 & \({\scriptscriptstyle ^0}\) & 1 & \({\scriptscriptstyle ^0}\) & \({\scriptscriptstyle ^0}\) & \({\scriptscriptstyle ^0}\) & \({\scriptscriptstyle ^0}\) & \({\scriptscriptstyle ^0}\) & \({\scriptscriptstyle ^0}\) & 1 & \({\scriptscriptstyle ^0}\) & 1 \\
 \(\!e_7\!\) & \({\scriptscriptstyle ^0}\) & \({\scriptscriptstyle ^0}\) & \({\scriptscriptstyle ^0}\) & \({\scriptscriptstyle ^0}\) & \({\scriptscriptstyle ^0}\) & 1 & \({\scriptscriptstyle ^0}\) & 1 & \({\scriptscriptstyle ^0}\) & \({\scriptscriptstyle ^0}\) & \({\scriptscriptstyle ^0}\) & \({\scriptscriptstyle ^0}\) & 1 & \({\scriptscriptstyle ^0}\) & 1 & \({\scriptscriptstyle ^0}\) \\
 \(\!e_8\!\) & \({\scriptscriptstyle ^0}\) & \({\scriptscriptstyle ^0}\) & \({\scriptscriptstyle ^0}\) & \({\scriptscriptstyle ^0}\) & \({\scriptscriptstyle ^0}\) & \({\scriptscriptstyle ^0}\) & \({\scriptscriptstyle ^0}\) & \({\scriptscriptstyle ^0}\) & 1 & 1 & \({\scriptscriptstyle ^0}\) & \({\scriptscriptstyle ^0}\) & \({\scriptscriptstyle ^0}\) & \({\scriptscriptstyle ^0}\) & 1 & 1 
\end{tabular}
\end{align*}
whereas the second scenario is the given by the Foulis-Randall product of the two Bell scenarios ${H_2\defin\mathcal{B}_{1,2,2}{^{\FR}\otimes}\,\mathcal{B}_{1,2,2}}$. Importantly the second scenario contains the hyperedges of the first, but has four additional hyperedges as well. Explicitly, $E(H_2)=E(H_1)\cup E(\Delta)$, where $E(\Delta)\defin$
\begin{align*}
\begin{tabular}{l|cccccccccccccccc}
  & 
\(\!\!\bigl(\!\begin{smallmatrix}
 0|0 \\
 0|0 \\
\end{smallmatrix}\!\bigr)\!\!\)
 & 
\(\!\!\bigl(\!\begin{smallmatrix}
 0|0 \\
 1|0 \\
\end{smallmatrix}\!\bigr)\!\!\)
 & 
\(\!\!\bigl(\!\begin{smallmatrix}
 1|0 \\
 0|0 \\
\end{smallmatrix}\!\bigr)\!\!\)
 & 
\(\!\!\bigl(\!\begin{smallmatrix}
 1|0 \\
 1|0 \\
\end{smallmatrix}\!\bigr)\!\!\)
 & 
\(\!\!\bigl(\!\begin{smallmatrix}
 0|0 \\
 0|1 \\
\end{smallmatrix}\!\bigr)\!\!\)
 & 
\(\!\!\bigl(\!\begin{smallmatrix}
 0|0 \\
 1|1 \\
\end{smallmatrix}\!\bigr)\!\!\)
 & 
\(\!\!\bigl(\!\begin{smallmatrix}
 1|0 \\
 0|1 \\
\end{smallmatrix}\!\bigr)\!\!\)
 & 
\(\!\!\bigl(\!\begin{smallmatrix}
 1|0 \\
 1|1 \\
\end{smallmatrix}\!\bigr)\!\!\)
 & 
\(\!\!\bigl(\!\begin{smallmatrix}
 0|1 \\
 0|0 \\
\end{smallmatrix}\!\bigr)\!\!\)
 & 
\(\!\!\bigl(\!\begin{smallmatrix}
 0|1 \\
 1|0 \\
\end{smallmatrix}\!\bigr)\!\!\)
 & 
\(\!\!\bigl(\!\begin{smallmatrix}
 1|1 \\
 0|0 \\
\end{smallmatrix}\!\bigr)\!\!\)
 & 
\(\!\!\bigl(\!\begin{smallmatrix}
 1|1 \\
 1|0 \\
\end{smallmatrix}\!\bigr)\!\!\)
 & 
\(\!\!\bigl(\!\begin{smallmatrix}
 0|1 \\
 0|1 \\
\end{smallmatrix}\!\bigr)\!\!\)
 & 
\(\!\!\bigl(\!\begin{smallmatrix}
 0|1 \\
 1|1 \\
\end{smallmatrix}\!\bigr)\!\!\)
 & 
\(\!\!\bigl(\!\begin{smallmatrix}
 1|1 \\
 0|1 \\
\end{smallmatrix}\!\bigr)\!\!\)
 & 
\(\!\!\bigl(\!\begin{smallmatrix}
 1|1 \\
 1|1 \\
\end{smallmatrix}\!\bigr)\!\!\)
 \\\hline
 \(\!e_9\!\) & \({\scriptscriptstyle ^0}\) & \({\scriptscriptstyle ^0}\) & \({\scriptscriptstyle ^0}\) & \({\scriptscriptstyle ^0}\) & \({\scriptscriptstyle ^0}\) & \({\scriptscriptstyle ^0}\) & \({\scriptscriptstyle ^0}\) & \({\scriptscriptstyle ^0}\) & \({\scriptscriptstyle ^0}\) & \({\scriptscriptstyle ^0}\) & \({\scriptscriptstyle ^0}\) & \({\scriptscriptstyle ^0}\) & 1 & 1 & 1 & 1 \\
  \(\!e_{10}\!\) & \({\scriptscriptstyle ^0}\) & \({\scriptscriptstyle ^0}\) & \({\scriptscriptstyle ^0}\) & \({\scriptscriptstyle ^0}\) & \({\scriptscriptstyle ^0}\) & \({\scriptscriptstyle ^0}\) & \({\scriptscriptstyle ^0}\) & \({\scriptscriptstyle ^0}\) & \({\scriptscriptstyle ^0}\) & \({\scriptscriptstyle ^0}\) & 1 & 1 & 1 & 1 & \({\scriptscriptstyle ^0}\) & \({\scriptscriptstyle ^0}\) \\
  \(\!e_{11}\!\) & \({\scriptscriptstyle ^0}\) & \({\scriptscriptstyle ^0}\) & \({\scriptscriptstyle ^0}\) & \({\scriptscriptstyle ^0}\) & \({\scriptscriptstyle ^0}\) & \({\scriptscriptstyle ^0}\) & \({\scriptscriptstyle ^0}\) & \({\scriptscriptstyle ^0}\) & 1 & 1 & 1 & 1 & \({\scriptscriptstyle ^0}\) & \({\scriptscriptstyle ^0}\) & \({\scriptscriptstyle ^0}\) & \({\scriptscriptstyle ^0}\) \\
  \(\!e_{12}\!\) & \({\scriptscriptstyle ^0}\) & \({\scriptscriptstyle ^0}\) & \({\scriptscriptstyle ^0}\) & \({\scriptscriptstyle ^0}\) & 1 & 1 & 1 & 1 & \({\scriptscriptstyle ^0}\) & \({\scriptscriptstyle ^0}\) & \({\scriptscriptstyle ^0}\) & \({\scriptscriptstyle ^0}\) & \({\scriptscriptstyle ^0}\) & \({\scriptscriptstyle ^0}\) & \({\scriptscriptstyle ^0}\) & \({\scriptscriptstyle ^0}\)
\end{tabular}
\end{align*}
\begin{sloppypar}
From these definitions, one can check that $H_1$ and $H_2$ have the same completion since $\mathtt{rref}(E^{\vec{\bf{1}}}(H_1))=\mathtt{rref}(E^{\vec{\bf{1}}}(H_2))$. In other words, each of the rows $e_9$-$e_{12}$ can be expressed as a linear combination of the rows $e_1$-$e_8$ using only $\pm 1$ coefficients. Indeed,  $e_9$-$e_{12}$ are virtual hyperedges of $H_1$.
\end{sloppypar}

Therefore, for the sake of the example take $H=H_1$, whose completion is $\overbar{H} = H_2$.  Consider now the Bell inequality given by \cite{AQ, JV}
\begin{align}\label{eq:ineqAQ}
\tfrac{30}{31} p_A(0|0) - \tfrac{167}{9} p_A(0|1) -\tfrac{167}{9} p_B(0|0) + \tfrac{30}{31} p_B(0|1) + \tfrac{174}{11} p_{AB}(00|00) \\ \nonumber
+ \tfrac{244}{23} p_{AB}(00|10) -\tfrac{74}{11} p_{AB}(00|01) + \tfrac{174}{11} p_{AB} (00|11) \leq 0.9677\,.
\end{align} 
Correlations in $\cQ_1(\overbar{H})$ achieve a maximum value for the left hand side of Ineq.~\eqref{eq:ineqAQ} of $1.02325$, whereas those in $\cQ_1(H)$ reach $1.15324$. Therefore $\cQ_1(\overbar{H}) \subsetneq \cQ_1(H)$, i.e. the completion scenario admits strictly fewer  $\cQ_1$ models than the incomplete scenario.

\end{document}